\documentclass[prd,longbibliography,nofootinbib,twocolumn,preprintnumbers]{revtex4-1}
\usepackage{amsmath,amssymb,amsthm}
\usepackage{graphicx}
\usepackage[usenames, dvipsnames]{xcolor}
\usepackage{slashed}
\usepackage{subfig}
\usepackage{tabularx}
\usepackage{empheq}
\usepackage{mathrsfs}
\usepackage{comment}
\usepackage{bbold}
\usepackage[shortlabels]{enumitem}
\usepackage[normalem]{ulem}
\usepackage{url}
\usepackage{color}

\definecolor{light-gray}{gray}{0.9}
\definecolor{medium-gray}{gray}{0.7}
\definecolor{darkblue}{rgb}{0.0,0.0,0.6}
\definecolor{red}{rgb}{0.9, 0,0}
\definecolor{navy}{rgb}{0.05, 0.05,0.8}
\definecolor{linkcolor}{rgb}{0.0, 0.28, 0.67}
\definecolor{paleblue}{rgb}{0.69, 0.93, 0.93}  

\usepackage[colorlinks]{hyperref}
\hypersetup{
     colorlinks   = true,
     citecolor    = linkcolor,
     linkcolor    = linkcolor,
     urlcolor    = linkcolor
}
\newcommand{\be}{\begin{equation}}
\newcommand{\ee}{\end{equation}}
\newcommand{\nl}{\nonumber \\}
\newcommand{\x}{\chi}

\newcommand{\mAp}{m_{A^\prime}}

\newcommand{\eps}{\epsilon}

\newcommand{\p}{\prime}

\newcommand{\grad}{\nabla}

\newcommand{\order}[1]{\mathcal{O}{(#1)}}
\newcommand{\Eq}[1]{Eq.~\ref{eq:#1}}

\newcommand{\Eqs}[2]{Eqs.~\ref{eq:#1} and \ref{eq:#2}}

\newcommand{\Fig}[1]{Fig.~\ref{fig:#1}}
\newcommand{\Figs}[2]{Figs.~\ref{fig:#1} and \ref{fig:#2}}

\newcommand{\bebox}{\begin{empheq}[box=\fcolorbox{light-gray}{light-gray}]{align}}
\newcommand{\eebox}{\end{empheq}}

\newcommand{\DM}{{_\text{DM}}}

\newcommand{\dt}{\partial_t}

\newcommand{\w}{\omega}

\newcommand{\n}{\nu}
\newcommand{\g}{\gamma}

\newcommand{\jv}{\boldsymbol{j}}

\newcommand{\xv}{{\bf x}}

\newcommand{\vv}{{\bf v}}
\newcommand{\Vv}{{\bf V}}

\newcommand{\Ev}{{\bf E}}

\newcommand{\eV}{\text{eV}}
\newcommand{\meV}{\text{meV}}

\newcommand{\MeV}{\text{MeV}}
\newcommand{\GeV}{\text{GeV}}

\newcommand{\V}{\text{V}}
\newcommand{\muV}{\mu \text{V}}
\newcommand{\nV}{\text{nV}}
\newcommand{\kV}{\text{kV}}
\newcommand{\MV}{\text{MV}}

\newcommand{\cm}{\text{cm}}
\newcommand{\mm}{\text{mm}}

\newcommand{\hr}{\text{hr}}
\newcommand{\yr}{\text{yr}}

\newcommand{\MHz}{\text{MHz}}
\newcommand{\kHz}{\text{kHz}}
\newcommand{\Hz}{\text{Hz}}

\usepackage{soul}

\begin{document}

\preprint{FERMILAB-PUB-25-0623-SQMS-T}

\title{Cavendish Tests of Millicharged Particles}

\author{Asher Berlin$^{a,b}$}
\author{Zachary Bogorad$^{a,b}$}
\author{Peter W.~Graham$^{c,d}$}
\author{Harikrishnan Ramani$^{e}$}
\affiliation{$^a$Theory Division, Fermi National Accelerator Laboratory}
\affiliation{$^b$Superconducting Quantum Materials and Systems Center (SQMS), Fermi National Accelerator Laboratory}
\affiliation{$^c$Leinweber Institute for Theoretical Physics at Stanford, Department of Physics, Stanford University}
\affiliation{$^d$Kavli Institute for Particle Astrophysics \& Cosmology, Department of Physics, Stanford
University}
\affiliation{$^e$Department of Physics and Astronomy, University of Delaware and the Bartol Research Institute}

\begin{abstract}
A terrestrial population of room-temperature millicharged particles can arise if they make up a dark matter subcomponent or if they are light enough to be produced in cosmic ray air showers. In a companion paper, we showed that a simple electrified shell acts as an efficient accumulator for such particles, parametrically enhancing their local density by many orders of magnitude. Here we demonstrate that Cavendish tests of Coulomb's Law, performed since the late 18th century, function as both quasistatic accumulators and detectors for this overdensity. Reinterpretations of these past Cavendish tests thus provide some of the strongest bounds on a terrestrial millicharge population. We also propose surrounding a Cavendish test with an additional charged shell, which significantly improves the sensitivity and can even enable detection of the irreducible density of millicharged particles generated from cosmic rays. Using decades-old technology, this can outperform future accelerator searches for sub-GeV masses.
\end{abstract}

\maketitle

One of the simplest extensions to the Standard Model (SM) involves the introduction of new particles $\x$ with a small effective electromagnetic charge $e q_\x \ll 1$.  Over the years such millicharged particles (mCPs) have attracted much attention~\cite{PVLAS:2005sku,Chang:2008aa,PAMELA:2008gwm,Barkana:2018lgd,Berlin:2018sjs,Barkana:2018qrx,Liu:2019knx} and many searches have been dedicated to finding them.  Accelerator searches to produce and detect these particles~\cite{Davidson:2000hf,Haas:2014dda,Prinz:1998ua,ArgoNeuT:2019ckq,milliQan:2021lne,ArguellesDelgado:2021lek,PBC:2025sny,CMS:2024eyx,Alcott:2025rxn} and SN1987A~\cite{Chang:2018rso} have set limits on their existence. Additionally, ion trap~\cite{Budker:2021quh}, matter neutrality~\cite{Kim:2007zzs,Moore:2014yba,Afek:2020lek}, and dark matter (DM) direct detection~\cite{SENSEI:2023zdf,DAMIC-M:2025luv,Iles:2024zka} experiments have been used to search for a local population of mCPs. Despite their simplicity and decades of scrutiny, mCPs are surprisingly unconstrained, with viable parameter space remaining for charges as large as $q_\x \sim 10^{-1}$ for GeV-scale masses.

Throughout a large range of parameter space, mCPs rapidly scatter with terrestrial matter and thermalize down to room temperature, $300 \ \text{K} \sim 25 \ \meV$, resulting in large overdensities on Earth~\cite{Pospelov:2020ktu,Berlin:2023zpn}. In this work, we highlight a novel way to detect  such a population, which could arise if, e.g., mCPs make up a DM subcomponent or are produced locally in cosmic ray air showers~\cite{Plestid:2020kdm} or nuclear decays~\cite{Gao:2025ykc}. 

Here, we show that century-old Cavendish-type tests of Coulomb's law  can be reinterpreted to set some of the strongest limits on terrestrial mCPs. In Cavendish experiments, the electric field is measured inside an empty conducting shell, whose surface charge is driven by a 
quasistatic oscillating voltage source. Since mCPs are able to penetrate conducting surfaces, the electric field generated outside the shell induces a charge density of mCPs on the interior. 
A non-zero electric field measured inside the driven shell thus constitutes either a violation of Coulomb's law (which can arise from, e.g., a non-zero photon mass or virtual particles~\cite{Jaeckel:2009dh}) or the presence of a pervasive background of mCPs, the latter of which is the focus of this work. 

In a companion paper~\cite{forthcoming}, we demonstrated that a shell held at fixed voltage (e.g., a Van de Graaff generator) functions as an accumulator for terrestrial mCPs (regardless of their origin) by dragging them inwards where they scatter and become electrically trapped in the interior, enhancing their local density by as much as twelve orders of magnitude. In this work, we propose enclosing a Cavendish experiment within such an accumulator. This minimal modification allows for the exploration of an even larger region of new parameter space, with sensitivity to the irreducible mCP density generated from cosmic rays. Note that a search for this irreducible population is equivalent to testing the model itself, independent of the mCP relic abundance, akin to an accelerator search. As we show below, a simple Cavendish experiment modified in this way can outperform future accelerator searches for sub-GeV mCPs. 

\vspace{0.25cm}
\noindent \textbf{Cavendish Tests.}---Since the mid-18th century, various experiments have been conducted to search for small deviations from Coulomb's inverse square law~\cite{GSpavieri_2004,Tu:2005ge}. A well-known example, originally performed by Henry Cavendish in 1773~\cite{maxwell}, involves measuring an electric field inside of a large charged conducting shell. While no such field is present when Gauss's law is exact, violations of it will generally lead to a non-zero field in the interior of the shell. A typical implementation of such a ``Cavendish test" uses an isolated series of concentric shells, where an outer shell is quasistatically oscillated at a frequency $\n_0$ and voltage $\phi_0$ relative to ground, and an oscillating potential difference between two interior shells is measured~\cite{maxwell,Plimpton:1936ont,Bartlett:1970js,Cochran1968,Williams:1971ms,Crandall:1983ec}. The most sensitive of these was performed in Ref.~\cite{Williams:1971ms}, which bounded the interior voltage difference to be smaller than $\sim 10^{-12} \ \V$.

Since the 20th century, these have often been reinterpreted as measurements of the photon mass $m_\g$.  In particular, $m_\g \neq 0$ modifies Gauss's law to be $\grad \cdot \Ev = \rho - m_\g^2 \, \phi$~\cite{Tu:2005ge}, where $\rho$ is the local charge density and $\phi$ is the electric potential. In a Cavendish test, a photon mass is thus equivalent, from the perspective of the detector, to an effective uniform charge density $\rho_\text{eff} = -m_\g^2 \, \phi_0$ within the driven shell. In the absence of physical charges in the interior, $\rho = 0$, the electric field a distance $r$ from the center is then $E \simeq \rho_\text{eff} \, r / 3$, which oscillates at the same frequency as the driven voltage.

This test of the photon mass is also necessarily a search for physical charges that are electrically pulled within the shells. In principle, both mCPs and SM ions can contribute to this effect. Crucially, however, the presence of the shells shields against SM charges while having a more limited effect on mCPs. 
This allows for the effective exclusion of SM ions by the experiment's shells, while still allowing mCPs to diffuse into the interior.
In particular, the work function of a conductor makes a barrier to positive mCPs $\sim q_\chi \times \left( \text{few eV}  \right)$. As a result, room-temperature mCPs with charge $\lesssim 300 \ \text{K} / (\text{few} \, \eV) \sim 10^{-2}$ can pass freely through metallic surfaces. 

\vspace{0.25cm}
\noindent \textbf{Millicharge Signals.}---In a Cavendish experiment, a signal arises from the charge density $\rho_\x$ of mCPs distributed uniformly inside the electrically-charged shell of voltage $\phi_0$ and radius $R_0$. In the continuum limit where there are many mCPs within the shell, this sources a measurable potential difference between radii $r_1 , r_2 < R_0$ of the form
\be
\label{eq:rhosignal}
\Delta \phi_\x = ( \rho_\x / 6 ) \, (r_2^2 - r_1^2)
~.
\ee

The behavior of mCPs can be divided into two regimes, depending on the degree to which their phase space is perturbed by the electric potential $\phi_0 \cos \left( \n_0 t \right)$ of the driven shell. First, in the small-coupling regime $e q_\x \phi_0  \ll T_\x$,  the mCP's phase space is only slightly perturbed by the charged shell. Here, $T_\x$ is the mCP's ambient temperature, which is approximately equal to the environment's temperature throughout the experiment, due to rapid scattering with nuclei. In this case, the amplitude of the oscillating induced charge density follows from the standard result for Debye screening in a weakly-coupled plasma~\cite{LandauKin,Berlin:2019uco}, corresponding to
\be
\label{eq:Debye}
\rho_\x \simeq  \varepsilon_\text{weak} \, m_D^2 \, \phi_0
~,
\ee
where $m_D = e q_\x \, \sqrt{n_\x / T_\x}$ is the mCP's contribution to the photon's Debye mass. In \Eq{Debye}, we have also introduced the dimensionless efficiency factor  $\varepsilon_\text{weak} \leq 1$ that accounts for the possibility that mCPs do not traverse the whole experiment within one oscillation. We define this precisely below. This is derived in Eqs.~\ref{eq:DebyeBallistic}, \ref{eq:boltzmannfactors}, \ref{eq:selfcollisional}, and \ref{eq:rhocoll2} in the Supplemental Material for a collisional or free-streaming thermal distribution. Comparing \Eq{Debye} to $\rho_\text{eff}$ for a massive photon, we see that $\sqrt{\varepsilon_\text{weak}} \, m_D$ plays the role of a photon mass in terms of its contribution as a source term in Gauss's Law.

Next, for large couplings $q_\x \gg T_\x / (e \phi_0)$, the form of the induced charge density depends on whether mCPs lose energy from scattering within the charged shell. For instance, as shown in the Supplemental Material, for free-streaming ballistic particles the charge density is smaller than the weak-coupling form in \Eq{Debye} by a factor of $\sim \sqrt{T_\x / e q_\x \phi_0} \ll 1$. However, $\rho_\x$ can instead be greatly enhanced if mCPs are able to efficiently diffuse into the charged shell within an oscillation time of the voltage,  lose energy via scattering (with, e.g., the inner wall or air), and become electrically bound to the electrostatic potential until it flips sign. Thus, the net charge density inside the shell still oscillates with frequency $\n_0$. As derived in Ref.~\cite{forthcoming}, this accumulated charge density over a  half-oscillation-time $t_\text{osc} = \n_0^{-1} / 2$ of the voltage is
\be
\label{eq:rhoxStrong}
\rho_\x \simeq  \frac{2}{\pi} \, \varepsilon_\text{strong} \, \frac{3  \, e q_\x  n_\x  \, V_E  \, t_\text{osc}}{R_0}
~,
\ee
where we have included a factor of $2 / \pi$ from time-averaging over a half oscillation. Since the above expression assumes a continuous distribution of mCPs, in our analysis we will restrict ourselves to a total accumulated charge larger than $(\rho_\x / e q_\x) \, R_0^3 \gtrsim 10^2$. In \Eq{rhoxStrong}, $\varepsilon_\text{strong} \leq 1$ is a dimensionless efficiency factor that is $\order{1}$ for an endless supply of mCPs that both diffuse and collisionally thermalize into the shell within an oscillation time (this is discussed more below as well as in Ref.~\cite{forthcoming}). The effective accumulation velocity $V_E$ is given by~\cite{forthcoming}, 
\be
\label{eq:VE}
V_E \simeq \frac{e q_\x \, E_0 / m_\x}{\max \big( \Gamma_p^\text{(air)} \, \beta_E \, , \, v_\text{th} / 2 R_0 \big) } \, 
~,
\ee
where $E_0 = \phi_0 / R_0$ is the electric field of the shell at its surface, $\Gamma_p^\text{(air)}$ is the momentum-exchange rate for mCP-atomic collisions in air (see, e.g., Refs.~\cite{Berlin:2023zpn,forthcoming}), and $v_\text{th} \simeq \sqrt{3 T_\x / m_\x}$ is the mCP thermal velocity. 

In \Eq{VE}, we have introduced the dimensionless parameter $\beta_E$, which accounts for modifications of the electric field due to image charges from nearby conductors. As derived in Ref.~\cite{forthcoming}, for an electric field that scales as a monopole or dipole at far distances, $\beta_E \simeq 1$ or $\beta_E \simeq \sqrt{\pi e q_\x \phi_0 / (2 T_\x)} \gg 1$, respectively. This is relevant in regards to whether the experiment is placed indoors or outdoors; for an outdoor experiment, Earth's crust generates an image charge which screens the monopole field at far distances~\cite{forthcoming}, whereas for an indoor experiment we take the field to be a monopole field at short distances and then to be zero outside the (assumed conducting) walls of the room.

As an example, let us consider an indoor experiment, such that the electric field is monopole-like wherever it is non-zero, corresponding to $\beta_E \simeq 1$. In this case, \Eq{rhoxStrong} can be rewritten in a suggestive form, 
\be
\label{eq:rhoxStrong2}
\rho_\x \simeq  \varepsilon_\text{strong} \, m_D^2 \phi_0 ~ \frac{6 \, t_\text{osc}}{\max{(t_\text{diff} \, , \, t_\text{th})}}
~,
\ee
where $t_\text{diff} = \pi R_0^2 / D_\x^\text{(air)}$ and $t_\text{th} = 3 \pi R_0 / (2 v_\text{th})$ are roughly the time it takes an mCP to collisionally-diffuse or thermally free-stream a distance $R_0$, respectively, and $D_\x^\text{(air)} = T_\x / (m_\x \, \Gamma_p^\text{(air)})$ is the mCP thermal diffusion coefficient in air~\cite{Berlin:2023zpn}. Compared to \Eq{Debye}, \Eq{rhoxStrong2} is thus enhanced by the ratio of the oscillation time $t_\text{osc}$ and the mCP crossing time.

The efficiency factors $\varepsilon_\text{weak}$ and $\varepsilon_\text{strong}$ of \Eqs{Debye}{rhoxStrong} can be decomposed further as
\be
\label{eq:epsdecompose}
\varepsilon_\text{weak} = \varepsilon_\text{diff} \, \varepsilon_g
~~,~~
\varepsilon_\text{strong} = \varepsilon_\text{diff} \, \varepsilon_g \, \varepsilon_\text{room} \, \varepsilon_\text{bound}
~.
\ee
The first factor $\varepsilon_\text{diff}$ accounts for the finite time it takes for mCPs to diffuse throughout the experimental volume. As derived in the Supplemental Material, it is given by
\be
\label{eq:diff}
\varepsilon_\text{diff} = \text{exp} \bigg( \hspace{-0.1 cm} - \max \bigg[ \bigg( \frac{t_\text{diff}}{2 t_\text{osc}} \bigg)^{1/2} , \, \bigg( \frac{3 t_\text{th}}{4 t_\text{osc}} \bigg)^{2/3} \, \bigg] \, \bigg)
~.
\ee
Thus, $\rho_\x$ is exponentially suppressed if the time $t_\text{diff} + t_\text{th}$ for mCPs to cross the experimental volume is large compared to the oscillation time $t_\text{osc}$. The other factors in \Eq{epsdecompose} are discussed in detail in Ref.~\cite{forthcoming}:  $\varepsilon_g$ accounts for the suppression of the shell's ability to attract mCPs if it  cannot overcome Earth's gravitational field; $\varepsilon_\text{room}$ incorporates the fact that accumulation of mCPs is limited by diffusion through the walls of the enclosing room if the shell is placed indoors; and $\varepsilon_\text{bound}$ accounts for the ability of mCPs to efficiently exchange energy and become electrically bound to the potential of the shell through scattering with atoms in the interior. 

As discussed in detail in Ref.~\cite{forthcoming}, the form of \Eq{rhoxStrong} neglects backreactions from coherent mCP self-interactions. This can arise if the repulsive force from the accumulated mCPs significantly perturbs the phase-space of newly arriving particles. This is most relevant when the interactions are mediated by a kinetically-mixed dark photon~\cite{Holdom:1985ag} and the mCP's contribution to the dark photon's Debye mass is larger than the inverse size of the shell, $m_D^\p \gtrsim R_0^{-1}$. For the Cavendish experiments discussed below, such effects are negligible if the accumulated charge density is smaller than $\rho_\x \lesssim 10^{9} \ \cm^{-3} \times (e q_\x / \alpha^\p)$, where $\alpha^\p$ is the dark photon fine-structure constant. Since this corresponds to charge densities well above the sensitivity thresholds of the detectors discussed below, this does not significantly impact our final results.

\vspace{0.25cm}
\noindent \textbf{Recasting Cavendish Tests.}---\Eq{diff} implies that it is advantageous for a Cavendish experiment to oscillate the voltage at a sufficiently low frequency, since this allows the mCPs ample time to pass throughout the experimental volume. However, a variety of systematic noise sources are relevant for $\n_0 \lesssim 1 \ \Hz$. This is discussed in the 1936 work of Plimpton and Lawton~\cite{Plimpton:1936ont}, which attempted to perform a Cavendish test with a DC outer shell voltage, but found that this was limited by mV-scale noise attributed to contact potentials. This was overcome in the same work by instead operating at Hz-scale frequencies, where they were instead limited by $\sim \muV$ Johnson noise. 

Various other Cavendish tests were performed in the late 1960s and early 1970s, operating with reduced noise at higher frequencies of $\sim 100 \ \Hz$~\cite{Bartlett:1970js}, $\sim 1 \  \kHz$~\cite{Cochran1968}, and $\sim 1 \ \MHz$~\cite{Williams:1971ms}. Although Ref.~\cite{Williams:1971ms} obtained strong sensitivity to $\sim 10^{-12} \ \V$ signals, $\nu_0 \sim \MHz$ is much too large to have facilitated sensitivity to slowly moving mCPs. We will therefore focus specifically on recasting the work of Plimpton and Lawton (PL) from 1936~\cite{Plimpton:1936ont} and Bartlett, Goldhagen, and Phillips (BGP) from 1970~\cite{Bartlett:1970js}, since these searches operated with low noise ($\sim \muV$ and $\sim \nV$, respectively) and sufficiently slow frequencies to be sensitive to a terrestrial mCP population.

In Ref.~\cite{Plimpton:1936ont}, PL employed two shells with radii of $0.76 \ \text{m}$ and $0.61 \ \text{m}$. The outer shell was either kept static or driven at a frequency of $\n_0 \simeq 2.2 \ \Hz$ with a voltage amplitude of $\phi_0 \simeq 3 \ \kV$. The enclosed galvanometer did not show voltages in excess of $\sim 1 \ \text{mV}$ or $\sim 1 \ \muV$ for $\n_0 \ll 1 \ \Hz$ or $\n_0 \sim 2 \ \Hz$, respectively. Note that $\sim 1 \ \muV$ is consistent with the general expectation of thermal fluctuations between the shells, which have a mutual capacitance of $C \simeq 0.3 \ \text{nF}$ and thus voltage fluctuations at the level of $\sqrt{300 \ \text{K} / C} \sim \muV$. 

In recasting the results of PL, we adopt $\Delta \phi_\x \lesssim 1 \ \text{mV}$ for $\n_0 = 10^{-1} \ \Hz$ and $\Delta \phi_\x \lesssim 1 \ \muV$ for $\n_0 = 2.2 \ \Hz$ when setting limits on the mCP-induced potential difference. This latter noise level is equivalent to the voltage induced by a total charge of $Q \sim 10^3 \, e$ oscillating in and out of the setup every second, which can arise from a permeating background of mCPs. 

The one remaining parameter of PL's apparatus needed for our analysis is the thickness of the shells. This is unfortunately unknown (they are described only as ``sheet iron''), so we adopt a typical value of $1 \ \mm$ for our analysis. While this  introduces some uncertainty into our analysis, we have checked that varying the thickness between $1 \ \mm - 3 \ \cm$ does not significantly impact our final results. 

In Ref.~\cite{Bartlett:1970js}, BGP further developed the setup of PL by employing a series of five concentric aluminum, steel, and copper spheres. The outermost sphere of radius $1.48 \ \text{m}$ served as a local ground, and a voltage was applied to the next inner sphere of radius $0.55 \ \text{m}$. A lock-in amplifier was then used to measure the voltage difference between the two innermost spheres of radii $0.46 \ \text{m}$ and $0.38 \ \text{m}$. Finally, a middle sphere of radius $0.5 \ \text{m}$ was used as a shield. The total material thickness of the shells was $\sim 5 \ \cm$. 

BGP performed measurements at two sets of frequencies and voltages. Although a higher frequency run at $\n_0 \simeq 2.5 \ \kHz$ was able to obtain moderately smaller noise levels, of more interest to us is a lower frequency run at $\n_0 \simeq 250 \ \Hz$ that employed a voltage of $\phi_0 \simeq 70 \ \kV$. The average of eight such runs obtained a noise level of $1.2 \ \nV$, which we take as the limit on $\Delta \phi_\x$. Note that this is also consistent with thermal Johnson noise, since the phase-sensitive voltage measurement enabled by the lock-in amplifier narrows the measurement bandwidth to $\sim 1/t_\text{int}$, where $t_\text{int} \sim 1 \ \hr$ is the total integration time. As a result, for $2 \pi \, \n_0 \tau \gg 1$ (where $\tau = R C$ and $R \sim 10^8 \ \Omega$ is the amplifier's input resistance), thermal voltage fluctuations are suppressed to $\sqrt{300 \ \text{K}/C} \, / \sqrt{\n_0^2 t_\text{int} \tau} \sim \text{nV}$.\footnote{Note that this scaling with integration time is possible since the signal has a well-defined frequency set by $\n_0$.} 

Armed with these experimental inputs, recasting  past limits from PL~\cite{Plimpton:1936ont} and BGP~\cite{Bartlett:1970js} into the mCP parameter space is straightforward. For any assumed mCP mass and charge, the procedure described in Eqs.~\ref{eq:rhosignal} -- \ref{eq:diff} gives a prediction for the resulting charge density and induced voltage difference $\Delta \phi_\x$. Since both experiments were performed indoors, we take the electric field of the shell to be monopole-like, corresponding to $\beta_E \simeq 1$ in \Eq{VE}. 

As mentioned above, the fact that these experiments were operated indoors can affect their ultimate sensitivity. In particular, the size of the grounded structures enclosing the driven shell dictates the range of the electric field and thus limits the total number of accumulated mCPs. Since PL oscillated the voltage of the outermost shell, we take the enclosing room to be of radius $3 \ \text{m}$, with conducting walls of thickness $30 \ \cm$ and a vent of area $1 \ \text{m}^2$ with wind speed $10 \ \text{m} / \text{s}$. We have checked that this choice only introduces an $\order{1}$ uncertainty into the derived limits. BGP instead operated the voltage on an inner shell, and so we take the size of their enclosing structure to be that of the outermost shell. 

\begin{figure}[t!]
\centering
\includegraphics[width=0.5 \textwidth]{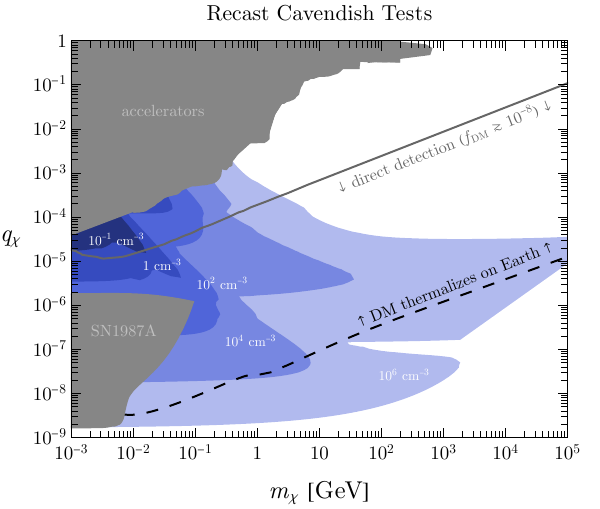}
\caption{New limits (blue) on the terrestrial density $n_\x$ of room-temperature millicharged particles (regardless of their origin), recast from past Cavendish experiments~\cite{Plimpton:1936ont,Bartlett:1970js}. Also shown are previous limits from accelerator probes~\cite{Davidson:2000hf,Haas:2014dda,Prinz:1998ua,ArgoNeuT:2019ckq,milliQan:2021lne,ArguellesDelgado:2021lek,PBC:2025sny,CMS:2024eyx,Alcott:2025rxn} and SN1987A~\cite{Chang:2018rso} (gray).  Above the solid gray line, millicharged dark matter quickly sheds its kinetic energy in Earth's atmosphere before reaching surface-level and underground direct detection experiments. Below this line, direct detection experiments are sensitive to a millicharged dark matter population, provided that these particles make up a sufficiently large fraction of the dark matter density, $f_\DM \gtrsim 10^{-8}$~\cite{Pospelov:2020ktu}. Above the dashed black line, millicharged dark matter is able to efficiently thermalize as it passes through Earth, which leads to large terrestrial overdensities (see Ref.~\cite{forthcoming}). A comparison to recent limits derived from ion trap experiments~\cite{Budker:2021quh} is shown in \Fig{ion}. }
\label{fig:recast}
\end{figure}

Our results are presented in \Fig{recast}, which shows the combined limits recast from PL and BGP for various choices of the ambient number density $n_\x$ (the individual limit derived from each experimental run is shown in the Supplemental Material in \Fig{CavSummary}). The various features are simple to understand at the qualitative level. First, neither PL nor BGP provides any meaningful sensitivity for sufficiently large couplings, since in this case the enhanced interaction strength implies that the time for mCPs to diffuse throughout the experiment is much longer than the voltage oscillation time. Limits derived from BGP appear at smaller couplings compared to those from PL, due to the larger operating frequency and smaller noise level. See the Supplemental Material for additional discussion along these lines.

For $m_\x \gtrsim q_\x \, E_0 / g_\oplus$ the gravitational field $g_\oplus$ of the Earth overcomes the electric field of the driven shell, preventing its ability to accumulate mCPs. Also note that the signal is suppressed for sufficiently small masses. This is due to the fact that the mCP-atomic scattering cross section is reduced at small momentum-transfer from shielding of the nuclear charge by atomic electrons. Furthermore, more collisions are required to efficiently exchange energy for mCPs much lighter than the nucleus. Both of these effects suppress the ability for MeV-scale mCPs to scatter and become bound to the charged shell. 

It is interesting to note that the limits shown in \Fig{recast} are competitive and complementary to those recently derived from ion traps~\cite{Carney:2021irt,Budker:2021quh}. As shown in \Fig{ion} in the Supplemental Material, these Cavendish tests are sensitive to smaller ambient number densities, as small as $n_\x \sim 10^{-1} \ \cm^{-3}$, as well as smaller couplings for the same assumed density. In comparison, current ion trap experiments are only sensitive to number densities of $n_\x \gtrsim 1 \ \cm^{-3}$ and to larger couplings. 

\vspace{0.25cm}
\noindent \textbf{Future Sensitivity with an Accumulator.}---As shown in our companion paper~\cite{forthcoming}, the local mCP density can be drastically enhanced inside an ``accumulator shell" held at fixed voltage. If a Cavendish experiment is operated in the interior, its sensitivity is enhanced by the larger density. In particular, we consider a setup with an outer accumulator shell of radius $R_\text{trap} = 2 \ \text{m}$ enclosing a ``Cavendish shell" of radius $R_0 = 1 \ \text{m}$. We take the Cavendish shell to  enclose a solid sphere of radius $0.5 \ \text{m}$, which functions to enhance the likelihood that mCPs scatter and lose energy within the experimental volume. For concreteness, we take all material to be made of iron, and the thickness of both shells is assumed to be $1  \ \mm$. The accumulator shell is assumed to be held at a fixed voltage of $\phi_\text{trap} = - 1 \ \text{MV}$ relative to ground; the sign of $\phi_\text{trap} < 0$ is chosen to target positively-charged mCPs, which remain unbound with atomic nuclei~\cite{forthcoming}, and which can accumulate due to Earth's electric field. 

The Cavendish shell and enclosed solid sphere are taken to operate similar to the Cavendish experiment of BGP from 1970~\cite{Bartlett:1970js}, corresponding to an oscillating shell potential of $\phi_0 = 50 \ \text{kV}$ with frequency $\n_0 = 250 \ \Hz$. The potential difference between the Cavendish shell and the sphere is assumed to be measured with a lock-in amplifier also comparable to that used by BGP, but with an extended integration time of $t_\text{int} = 1 \ \yr$ instead of $1 \ \hr$, corresponding to a reduced noise level of $10^{-9} \ \V \times \sqrt{\hr / \yr} \sim 10^{-11} \ \V$. In principle, these noise levels can be parametrically reduced through the use of a resonantly tuned high-$Q$ LC circuit. We do not consider this in detail, though, since these operate at cryogenic temperatures, substantially increasing the mCP diffusion time $t_\text{diff}$.

Even a relatively conservative implementation of this experimental concept that is operated indoors at room temperature and atmospheric pressure can parametrically enhance the sensitivity shown in \Fig{recast}. However, we will consider a few additional modifications in order to extend the discovery prospects further. For instance, in our estimates, the voltage of the accumulator shell is held fixed for a period of $t_\text{trap} = 1 \ \text{yr}$. Ideally, this should be operated outdoors, since the total number of trapped mCPs is ultimately limited by the size of an enclosure if it is placed indoors~\cite{forthcoming}. We also note that the time for an mCP to diffuse throughout several meters of room-temperature air becomes non-negligible compared to the voltage oscillation time for couplings $q_\x \gg 10^{-3}$. This motivates operating the inner region of the accumulator shell in vacuum, which significantly reduces the total diffusion time of the mCPs throughout the experimental volume. This requires maintaining vacuum through purely passive means, such as cryopumping, as most active measures would also deplete the trap of strongly-coupled mCPs.

In our estimates, we take the enclosed volume to operate at room temperature and at a pressure of $10^{-6} \ \text{atm}$ because this approximately saturates the benefit of operating with low vacuum (i.e., for smaller values, our results are approximately independent of this choice). Instead, for the largest mCP couplings we consider in this work, the reach of our setup is degraded for parametrically larger pressures, but is only mildly affected unless internal pressures exceed $10^{-4} \ \text{atm}$. In this low pressure environment, mCPs do not necessarily scatter with air molecules within the shells, in which case we approximate their motion as ballistic. This results in an increased kinetic energy as the mCPs are accelerated by the electric field of the inner shell, thereby further decreasing their diffusion time throughout the experiment and leading to an enhanced reach for larger charges. 

Following the formalism of Ref.~\cite{forthcoming}, we calculate the overdensity of mCPs accumulated by a trap placed outdoors, which can be used to determine the reach of this setup. The projected model-independent sensitivity is shown in the Supplemental Material in \Fig{future} for various choices of the terrestrial mCP density $n_\x$. Many of the qualitative features are similar to those in \Fig{recast}, with the main difference being that this setup is sensitive to significantly smaller number densities (as small as $n_\x \sim 10^{-16} \ \cm^{-3}$) and a much larger range of couplings. 

\begin{figure}[t]
\centering
\includegraphics[width=0.5 \textwidth]{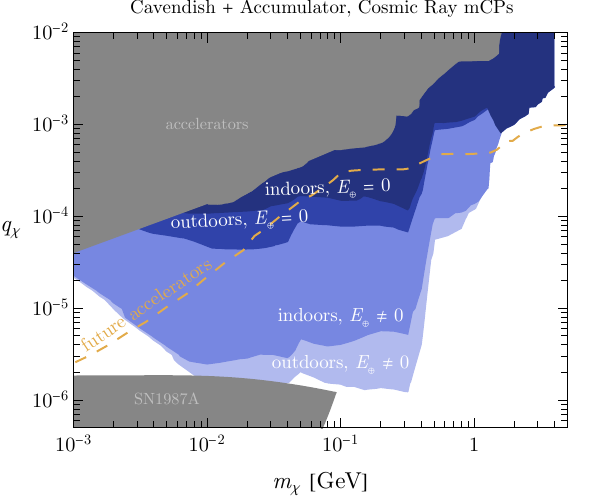}
\caption{The projected sensitivity of a dedicated Cavendish test enclosed in an electric trap to the irreducible terrestrial density of millicharged particles produced by cosmic rays. We take the millicharged particles to efficiently ($E_\oplus \neq 0$) or not efficiently ($E_\oplus = 0$) couple to the atmospheric electric field, which can substantially enhance the local density of particles near and below the crust. We show both possibilities for clarity, but as discussed in Ref.~\cite{forthcoming}, ignoring the effect of $E_\oplus$ is unrealistic, and thus a detailed modeling is likely to yield results much closer to our $E_\oplus \neq 0$ projections. This cosmic ray population is irreducible and solely a function of the charge and mass, such that testing this local density is equivalent to testing the model itself. We consider setups placed inside a room of radius $R_\text{room} = 10 \ \text{m}$ or placed outdoors. Also shown as the dashed orange line is a collection of projections for proposed future accelerator searches, taken from Refs.~\cite{Berlin:2018bsc,PBC:2025sny}.}
\label{fig:CRreach}
\end{figure}

One caveat can arise in kinetically-mixed models since mCPs effectively couple to both the dark and visible electric field. As a result, the accumulated mCPs can then source a dark electric field that dominates the effect of the SM fields, thus halting further accumulation. This occurs once $e^{\p \, 2}  n_\x R_0 \sim e q_\x E_0$, corresponding to an mCP-induced visible electric field $E_\x \sim e q_\x n_\x  R_0 \sim (e q_\x / e^\p)^2 E_0 \sim \eps^2 E_0$, where $\eps = e q_\x / e^\p$ is the kinetic mixing parameter and $e^\p$ is the dark photon gauge coupling. Hence, accumulation saturates once the mCPs screen an $\eps^2$ fraction of the driven Cavendish shell's visible field. This is thus irrelevant to our analysis for kinetic mixing parameters larger than $\eps \gtrsim \sqrt{\Delta \phi_\x / \phi_0} \sim \sqrt{1 \ \nV / 100 \ \kV} = 10^{-7}$. For the large majority of the parameter space explored in this paper, this model-dependence is not of concern. In particular, this leaves \Fig{CRreach} unaffected (to be discussed below), while for $e^\p \sim 1$ it is relevant for only a small range of couplings in \Figs{recast}{future}. We also note that mCP annihilations into dark photons can limit the size of the symmetric abundance; this is irrelevant except for the largest densities ($n_\x \sim 10^6 \ \cm^{-3}$ in \Fig{recast}) and smallest masses ($\sim \MeV$) that we consider, in which case such processes are important only for $e^\p \gg 10^{-2}$.

\vspace{0.25cm}
\noindent \textbf{Cosmic Ray Population.}---An irreducible population of mCPs is generated by decays of mesons in cosmic ray air showers. As shown in Ref.~\cite{forthcoming}, the resulting terrestrial density of thermalized mCPs can be as large as $n_\x \sim 10^{-5} \ \cm^{-3} \times (q_\x / 10^{-4})^2 \, $ for $m_\x \ll 1 \ \GeV$. We now use those results to show that a Cavendish experiment enclosed within an electrostatic accumulator is sensitive to this population. The projected reach is shown in \Fig{CRreach}, adopting the same experimental inputs as described above, and for both indoor and outdoor operations of the experiment. The indoor experiment is assumed to be surrounded by conducting walls. As described in Ref.~\cite{forthcoming}, it may be possible to operate an indoor setup with sensitivity comparable to that shown for an outdoor one, depending on the actual room design.

\Fig{CRreach} shows results assuming that positively-charged mCPs are either trapped or not trapped on Earth by the atmospheric electric field $E_\oplus \sim 1 \ \V / \cm$, which gives rise to an approximately stable $\phi_\oplus \sim 0.3 \ \MV$ voltage gap between the crust and ionosphere. As discussed below and in Ref.~\cite{Berlin:2023zpn}, mCPs necessarily couple to such long-ranged electromagnetic fields in this parameter space. Therefore, $\phi_\oplus$ is expected to prevent evaporation of light thermalized mCPs, significantly enhancing their local abundance. In \Fig{CRreach}, we have also presented the case where the role of the atmospheric electric field is neglected, mainly for the purpose of illustrating its enhancing effect; in this sense, the ``$E_\oplus = 0$" regions are unrealistically conservative, and we expect detailed modeling to yield results similar to or even stronger than our ``$E_\oplus \neq 0$" projections. This is because the local mCP density can be greatly increased in models involving  kinetically-mixed dark photon mediators, since the accumulation of positively-charged mCPs by $E_\oplus$ results in a dark electric field that forces mCPs to predominantly occupy a narrow radial region near Earth's surface (see Ref.~\cite{forthcoming} for additional discussion).

For comparison, we also show projections of various proposed accelerator searches, including LDMX~\cite{Berlin:2018bsc} and those at the LHC~\cite{PBC:2025sny}. From this we see that for sub-GeV masses, a Cavendish test (using existing technology and noise-levels achieved several decades ago) is competitive with, and for some masses exceeds, the ability of future accelerator searches.

In comparing Cavendish tests to accelerator searches, it is worthwhile to ask whether there are model-dependencies that can make this comparison less straightforward. For instance, unlike Cavendish tests, collider searches for mCPs do not fundamentally rely on the mCP-photon interaction being long-ranged on macroscopic length scales. This is always true for mCPs with a direct millicharge but is dependent on the dark photon mass for the kinetically-mixed model. In particular, the detection of accelerator-produced mCPs relies on low-threshold scintillation or ionization signals (see, e.g., Ref.~\cite{Magill:2018tbb,Harnik:2019zee,Oscura:2023qch,Essig:2024ebk}). As a result, in order to maximize the signal, such analyses implicitly assume that these interactions arise from a very light dark photon with mass $\mAp \lesssim \sqrt{m_e \, E_\text{th}} \lesssim 1 \ \MeV$, where $E_\text{th}$ is the detector threshold. In perturbative models, the fact that the kinetic mixing parameter is related to the millicharge by $\eps = e q_\x / e^\p$~\cite{Holdom:1985ag} implies that accelerator searches for $q_\x \gtrsim 10^{-5}$ inherently require $\eps \gtrsim 10^{-6}$. For such values of $\eps$, direct bounds on dark photons~\cite{Mirizzi:2009iz,An:2013yfc} imply that dark photons with mass below MeV must in fact have much lighter masses and be macroscopically long-ranged. Hence, both accelerators and Cavendish tests fundamentally explore the same model-space of mCPs with long-ranged interactions.

\vspace{0.25cm}
\noindent \textbf{Discussion.}---Cavendish experiments are incredibly sensitive to the small electric fields sourced by a low-density background of room-temperature millicharged particles. We have exemplified this by recasting century-old Cavendish tests to provide new bounds that are complementary, and sometimes stronger, than those derived from modern ion traps.

We have also proposed enclosing such an experiment within a static charged ``accumulator shell," analogous to a large Van de Graaff generator, which enhances the ambient millicharge density. Remarkably, this simple experimental setup employing decades-old technology can enable sensitivity to the irreducible terrestrial population sourced by cosmic rays, outperforming future accelerator searches for sub-GeV millicharged particles. Note that the inclusion of an accumulator shell also serves as a discovery tool; the millicharge origin of a signal could be confirmed by testing the correlation with runtime, size, and voltage of the accumulator. 

In this work, we have chosen to focus exclusively on Cavendish tests, due to their relative simplicity. However, note that various different kinds of sensors could be operated within an accumulator shell, such as ion traps, skipper CCDs, or other kinds of low-threshold sensors. We leave the exploration of such ideas to future work.

\vspace{0.25cm}
\section*{Acknowledgements}

We thank Paddy Fox and Kent Irwin for valuable conversations. This manuscript has been authored in part by Fermi Forward Discovery Group, LLC under Contract No. 89243024CSC000002 with the U.S. Department of Energy, Office of Science, Office of High Energy Physics. This material is based upon work supported by the U.S. Department of Energy, Office of Science, National Quantum Information Science Research Centers, Superconducting Quantum Materials and Systems Center (SQMS) under contract number DE-AC02-07CH11359. 
This work was supported in part by NSF Grant No.~PHY-2310429, NSF Grant No. PHY-2515007, Simons Investigator Award No.~824870,  the Gordon and Betty Moore Foundation Grant No.~GBMF7946, the
University of Delaware Research Foundation and the John Templeton Foundation Award No.~63595.

\bibliography{main}

\begin{thebibliography}{49}%
\makeatletter
\providecommand \@ifxundefined [1]{%
 \@ifx{#1\undefined}
}%
\providecommand \@ifnum [1]{%
 \ifnum #1\expandafter \@firstoftwo
 \else \expandafter \@secondoftwo
 \fi
}%
\providecommand \@ifx [1]{%
 \ifx #1\expandafter \@firstoftwo
 \else \expandafter \@secondoftwo
 \fi
}%
\providecommand \natexlab [1]{#1}%
\providecommand \enquote  [1]{``#1''}%
\providecommand \bibnamefont  [1]{#1}%
\providecommand \bibfnamefont [1]{#1}%
\providecommand \citenamefont [1]{#1}%
\providecommand \href@noop [0]{\@secondoftwo}%
\providecommand \href [0]{\begingroup \@sanitize@url \@href}%
\providecommand \@href[1]{\@@startlink{#1}\@@href}%
\providecommand \@@href[1]{\endgroup#1\@@endlink}%
\providecommand \@sanitize@url [0]{\catcode `\\12\catcode `\$12\catcode `\&12\catcode `\#12\catcode `\^12\catcode `\_12\catcode `\%12\relax}%
\providecommand \@@startlink[1]{}%
\providecommand \@@endlink[0]{}%
\providecommand \url  [0]{\begingroup\@sanitize@url \@url }%
\providecommand \@url [1]{\endgroup\@href {#1}{\urlprefix }}%
\providecommand \urlprefix  [0]{URL }%
\providecommand \Eprint [0]{\href }%
\providecommand \doibase [0]{http://dx.doi.org/}%
\providecommand \selectlanguage [0]{\@gobble}%
\providecommand \bibinfo  [0]{\@secondoftwo}%
\providecommand \bibfield  [0]{\@secondoftwo}%
\providecommand \translation [1]{[#1]}%
\providecommand \BibitemOpen [0]{}%
\providecommand \bibitemStop [0]{}%
\providecommand \bibitemNoStop [0]{.\EOS\space}%
\providecommand \EOS [0]{\spacefactor3000\relax}%
\providecommand \BibitemShut  [1]{\csname bibitem#1\endcsname}%
\let\auto@bib@innerbib\@empty
\bibitem [{\citenamefont {Zavattini}\ \emph {et~al.}(2006)\citenamefont {Zavattini} \emph {et~al.}}]{PVLAS:2005sku}%
  \BibitemOpen
  \bibfield  {author} {\bibinfo {author} {\bibfnamefont {E.}~\bibnamefont {Zavattini}} \emph {et~al.} (\bibinfo {collaboration} {PVLAS}),\ }\bibfield  {title} {\enquote {\bibinfo {title} {{Experimental observation of optical rotation generated in vacuum by a magnetic field}},}\ }\href {\doibase 10.1103/PhysRevLett.99.129901} {\bibfield  {journal} {\bibinfo  {journal} {Phys. Rev. Lett.}\ }\textbf {\bibinfo {volume} {96}},\ \bibinfo {pages} {110406} (\bibinfo {year} {2006})},\ \bibinfo {note} {[Erratum: Phys.Rev.Lett. 99, 129901 (2007)]},\ \Eprint {http://arxiv.org/abs/hep-ex/0507107} {arXiv:hep-ex/0507107} \BibitemShut {NoStop}%
\bibitem [{\citenamefont {Chang}\ \emph {et~al.}(2008)\citenamefont {Chang} \emph {et~al.}}]{Chang:2008aa}%
  \BibitemOpen
  \bibfield  {author} {\bibinfo {author} {\bibfnamefont {J.}~\bibnamefont {Chang}} \emph {et~al.},\ }\bibfield  {title} {\enquote {\bibinfo {title} {{An excess of cosmic ray electrons at energies of 300-800 GeV}},}\ }\href {\doibase 10.1038/nature07477} {\bibfield  {journal} {\bibinfo  {journal} {Nature}\ }\textbf {\bibinfo {volume} {456}},\ \bibinfo {pages} {362--365} (\bibinfo {year} {2008})}\BibitemShut {NoStop}%
\bibitem [{\citenamefont {Adriani}\ \emph {et~al.}(2009)\citenamefont {Adriani} \emph {et~al.}}]{PAMELA:2008gwm}%
  \BibitemOpen
  \bibfield  {author} {\bibinfo {author} {\bibfnamefont {Oscar}\ \bibnamefont {Adriani}} \emph {et~al.} (\bibinfo {collaboration} {PAMELA}),\ }\bibfield  {title} {\enquote {\bibinfo {title} {{An anomalous positron abundance in cosmic rays with energies 1.5-100 GeV}},}\ }\href {\doibase 10.1038/nature07942} {\bibfield  {journal} {\bibinfo  {journal} {Nature}\ }\textbf {\bibinfo {volume} {458}},\ \bibinfo {pages} {607--609} (\bibinfo {year} {2009})},\ \Eprint {http://arxiv.org/abs/0810.4995} {arXiv:0810.4995 [astro-ph]} \BibitemShut {NoStop}%
\bibitem [{\citenamefont {Barkana}(2018)}]{Barkana:2018lgd}%
  \BibitemOpen
  \bibfield  {author} {\bibinfo {author} {\bibfnamefont {Rennan}\ \bibnamefont {Barkana}},\ }\bibfield  {title} {\enquote {\bibinfo {title} {{Possible interaction between baryons and dark-matter particles revealed by the first stars}},}\ }\href {\doibase 10.1038/nature25791} {\bibfield  {journal} {\bibinfo  {journal} {Nature}\ }\textbf {\bibinfo {volume} {555}},\ \bibinfo {pages} {71--74} (\bibinfo {year} {2018})},\ \Eprint {http://arxiv.org/abs/1803.06698} {arXiv:1803.06698 [astro-ph.CO]} \BibitemShut {NoStop}%
\bibitem [{\citenamefont {Berlin}\ \emph {et~al.}(2018)\citenamefont {Berlin}, \citenamefont {Hooper}, \citenamefont {Krnjaic},\ and\ \citenamefont {McDermott}}]{Berlin:2018sjs}%
  \BibitemOpen
  \bibfield  {author} {\bibinfo {author} {\bibfnamefont {Asher}\ \bibnamefont {Berlin}}, \bibinfo {author} {\bibfnamefont {Dan}\ \bibnamefont {Hooper}}, \bibinfo {author} {\bibfnamefont {Gordan}\ \bibnamefont {Krnjaic}}, \ and\ \bibinfo {author} {\bibfnamefont {Samuel~D.}\ \bibnamefont {McDermott}},\ }\bibfield  {title} {\enquote {\bibinfo {title} {{Severely Constraining Dark Matter Interpretations of the 21-cm Anomaly}},}\ }\href {\doibase 10.1103/PhysRevLett.121.011102} {\bibfield  {journal} {\bibinfo  {journal} {Phys. Rev. Lett.}\ }\textbf {\bibinfo {volume} {121}},\ \bibinfo {pages} {011102} (\bibinfo {year} {2018})},\ \Eprint {http://arxiv.org/abs/1803.02804} {arXiv:1803.02804 [hep-ph]} \BibitemShut {NoStop}%
\bibitem [{\citenamefont {Barkana}\ \emph {et~al.}(2018)\citenamefont {Barkana}, \citenamefont {Outmezguine}, \citenamefont {Redigolo},\ and\ \citenamefont {Volansky}}]{Barkana:2018qrx}%
  \BibitemOpen
  \bibfield  {author} {\bibinfo {author} {\bibfnamefont {Rennan}\ \bibnamefont {Barkana}}, \bibinfo {author} {\bibfnamefont {Nadav~Joseph}\ \bibnamefont {Outmezguine}}, \bibinfo {author} {\bibfnamefont {Diego}\ \bibnamefont {Redigolo}}, \ and\ \bibinfo {author} {\bibfnamefont {Tomer}\ \bibnamefont {Volansky}},\ }\bibfield  {title} {\enquote {\bibinfo {title} {{Strong constraints on light dark matter interpretation of the EDGES signal}},}\ }\href {\doibase 10.1103/PhysRevD.98.103005} {\bibfield  {journal} {\bibinfo  {journal} {Phys. Rev. D}\ }\textbf {\bibinfo {volume} {98}},\ \bibinfo {pages} {103005} (\bibinfo {year} {2018})},\ \Eprint {http://arxiv.org/abs/1803.03091} {arXiv:1803.03091 [hep-ph]} \BibitemShut {NoStop}%
\bibitem [{\citenamefont {Liu}\ \emph {et~al.}(2019)\citenamefont {Liu}, \citenamefont {Outmezguine}, \citenamefont {Redigolo},\ and\ \citenamefont {Volansky}}]{Liu:2019knx}%
  \BibitemOpen
  \bibfield  {author} {\bibinfo {author} {\bibfnamefont {Hongwan}\ \bibnamefont {Liu}}, \bibinfo {author} {\bibfnamefont {Nadav~Joseph}\ \bibnamefont {Outmezguine}}, \bibinfo {author} {\bibfnamefont {Diego}\ \bibnamefont {Redigolo}}, \ and\ \bibinfo {author} {\bibfnamefont {Tomer}\ \bibnamefont {Volansky}},\ }\bibfield  {title} {\enquote {\bibinfo {title} {{Reviving Millicharged Dark Matter for 21-cm Cosmology}},}\ }\href {\doibase 10.1103/PhysRevD.100.123011} {\bibfield  {journal} {\bibinfo  {journal} {Phys. Rev. D}\ }\textbf {\bibinfo {volume} {100}},\ \bibinfo {pages} {123011} (\bibinfo {year} {2019})},\ \Eprint {http://arxiv.org/abs/1908.06986} {arXiv:1908.06986 [hep-ph]} \BibitemShut {NoStop}%
\bibitem [{\citenamefont {Davidson}\ \emph {et~al.}(2000)\citenamefont {Davidson}, \citenamefont {Hannestad},\ and\ \citenamefont {Raffelt}}]{Davidson:2000hf}%
  \BibitemOpen
  \bibfield  {author} {\bibinfo {author} {\bibfnamefont {Sacha}\ \bibnamefont {Davidson}}, \bibinfo {author} {\bibfnamefont {Steen}\ \bibnamefont {Hannestad}}, \ and\ \bibinfo {author} {\bibfnamefont {Georg}\ \bibnamefont {Raffelt}},\ }\bibfield  {title} {\enquote {\bibinfo {title} {{Updated bounds on millicharged particles}},}\ }\href {\doibase 10.1088/1126-6708/2000/05/003} {\bibfield  {journal} {\bibinfo  {journal} {JHEP}\ }\textbf {\bibinfo {volume} {05}},\ \bibinfo {pages} {003} (\bibinfo {year} {2000})},\ \Eprint {http://arxiv.org/abs/hep-ph/0001179} {arXiv:hep-ph/0001179} \BibitemShut {NoStop}%
\bibitem [{\citenamefont {Haas}\ \emph {et~al.}(2015)\citenamefont {Haas}, \citenamefont {Hill}, \citenamefont {Izaguirre},\ and\ \citenamefont {Yavin}}]{Haas:2014dda}%
  \BibitemOpen
  \bibfield  {author} {\bibinfo {author} {\bibfnamefont {Andrew}\ \bibnamefont {Haas}}, \bibinfo {author} {\bibfnamefont {Christopher~S.}\ \bibnamefont {Hill}}, \bibinfo {author} {\bibfnamefont {Eder}\ \bibnamefont {Izaguirre}}, \ and\ \bibinfo {author} {\bibfnamefont {Itay}\ \bibnamefont {Yavin}},\ }\bibfield  {title} {\enquote {\bibinfo {title} {{Looking for milli-charged particles with a new experiment at the LHC}},}\ }\href {\doibase 10.1016/j.physletb.2015.04.062} {\bibfield  {journal} {\bibinfo  {journal} {Phys. Lett. B}\ }\textbf {\bibinfo {volume} {746}},\ \bibinfo {pages} {117--120} (\bibinfo {year} {2015})},\ \Eprint {http://arxiv.org/abs/1410.6816} {arXiv:1410.6816 [hep-ph]} \BibitemShut {NoStop}%
\bibitem [{\citenamefont {Prinz}\ \emph {et~al.}(1998)\citenamefont {Prinz} \emph {et~al.}}]{Prinz:1998ua}%
  \BibitemOpen
  \bibfield  {author} {\bibinfo {author} {\bibfnamefont {A.~A.}\ \bibnamefont {Prinz}} \emph {et~al.},\ }\bibfield  {title} {\enquote {\bibinfo {title} {{Search for millicharged particles at SLAC}},}\ }\href {\doibase 10.1103/PhysRevLett.81.1175} {\bibfield  {journal} {\bibinfo  {journal} {Phys. Rev. Lett.}\ }\textbf {\bibinfo {volume} {81}},\ \bibinfo {pages} {1175--1178} (\bibinfo {year} {1998})},\ \Eprint {http://arxiv.org/abs/hep-ex/9804008} {arXiv:hep-ex/9804008} \BibitemShut {NoStop}%
\bibitem [{\citenamefont {Acciarri}\ \emph {et~al.}(2020)\citenamefont {Acciarri} \emph {et~al.}}]{ArgoNeuT:2019ckq}%
  \BibitemOpen
  \bibfield  {author} {\bibinfo {author} {\bibfnamefont {R.}~\bibnamefont {Acciarri}} \emph {et~al.} (\bibinfo {collaboration} {ArgoNeuT}),\ }\bibfield  {title} {\enquote {\bibinfo {title} {{Improved Limits on Millicharged Particles Using the ArgoNeuT Experiment at Fermilab}},}\ }\href {\doibase 10.1103/PhysRevLett.124.131801} {\bibfield  {journal} {\bibinfo  {journal} {Phys. Rev. Lett.}\ }\textbf {\bibinfo {volume} {124}},\ \bibinfo {pages} {131801} (\bibinfo {year} {2020})},\ \Eprint {http://arxiv.org/abs/1911.07996} {arXiv:1911.07996 [hep-ex]} \BibitemShut {NoStop}%
\bibitem [{\citenamefont {Ball}\ \emph {et~al.}(2021)\citenamefont {Ball} \emph {et~al.}}]{milliQan:2021lne}%
  \BibitemOpen
  \bibfield  {author} {\bibinfo {author} {\bibfnamefont {A.}~\bibnamefont {Ball}} \emph {et~al.} (\bibinfo {collaboration} {milliQan}),\ }\bibfield  {title} {\enquote {\bibinfo {title} {{Sensitivity to millicharged particles in future proton-proton collisions at the LHC with the milliQan detector}},}\ }\href {\doibase 10.1103/PhysRevD.104.032002} {\bibfield  {journal} {\bibinfo  {journal} {Phys. Rev. D}\ }\textbf {\bibinfo {volume} {104}},\ \bibinfo {pages} {032002} (\bibinfo {year} {2021})},\ \Eprint {http://arxiv.org/abs/2104.07151} {arXiv:2104.07151 [hep-ex]} \BibitemShut {NoStop}%
\bibitem [{\citenamefont {Arg\"uelles~Delgado}\ \emph {et~al.}(2021)\citenamefont {Arg\"uelles~Delgado}, \citenamefont {Kelly},\ and\ \citenamefont {Mu\~noz Albornoz}}]{ArguellesDelgado:2021lek}%
  \BibitemOpen
  \bibfield  {author} {\bibinfo {author} {\bibfnamefont {Carlos~Alberto}\ \bibnamefont {Arg\"uelles~Delgado}}, \bibinfo {author} {\bibfnamefont {Kevin~James}\ \bibnamefont {Kelly}}, \ and\ \bibinfo {author} {\bibfnamefont {V\'\i{}ctor}\ \bibnamefont {Mu\~noz Albornoz}},\ }\bibfield  {title} {\enquote {\bibinfo {title} {{Millicharged particles from the heavens: single- and multiple-scattering signatures}},}\ }\href {\doibase 10.1007/JHEP11(2021)099} {\bibfield  {journal} {\bibinfo  {journal} {JHEP}\ }\textbf {\bibinfo {volume} {11}},\ \bibinfo {pages} {099} (\bibinfo {year} {2021})},\ \Eprint {http://arxiv.org/abs/2104.13924} {arXiv:2104.13924 [hep-ph]} \BibitemShut {NoStop}%
\bibitem [{\citenamefont {Alemany~Fern\'andez}\ \emph {et~al.}(2025)\citenamefont {Alemany~Fern\'andez} \emph {et~al.}}]{PBC:2025sny}%
  \BibitemOpen
  \bibfield  {author} {\bibinfo {author} {\bibfnamefont {R.}~\bibnamefont {Alemany~Fern\'andez}} \emph {et~al.} (\bibinfo {collaboration} {PBC}),\ }\bibfield  {title} {\enquote {\bibinfo {title} {{Summary Report of the Physics Beyond Colliders Study at CERN}},}\ }\href@noop {} {\  (\bibinfo {year} {2025})},\ \Eprint {http://arxiv.org/abs/2505.00947} {arXiv:2505.00947 [hep-ex]} \BibitemShut {NoStop}%
\bibitem [{\citenamefont {Hayrapetyan}\ \emph {et~al.}(2025)\citenamefont {Hayrapetyan} \emph {et~al.}}]{CMS:2024eyx}%
  \BibitemOpen
  \bibfield  {author} {\bibinfo {author} {\bibfnamefont {Aram}\ \bibnamefont {Hayrapetyan}} \emph {et~al.} (\bibinfo {collaboration} {CMS}),\ }\bibfield  {title} {\enquote {\bibinfo {title} {{Search for Fractionally Charged Particles in Proton-Proton Collisions at s=13{\,}{\,}TeV}},}\ }\href {\doibase 10.1103/PhysRevLett.134.131802} {\bibfield  {journal} {\bibinfo  {journal} {Phys. Rev. Lett.}\ }\textbf {\bibinfo {volume} {134}},\ \bibinfo {pages} {131802} (\bibinfo {year} {2025})},\ \Eprint {http://arxiv.org/abs/2402.09932} {arXiv:2402.09932 [hep-ex]} \BibitemShut {NoStop}%
\bibitem [{\citenamefont {Alcott}\ \emph {et~al.}(2025)\citenamefont {Alcott} \emph {et~al.}}]{Alcott:2025rxn}%
  \BibitemOpen
  \bibfield  {author} {\bibinfo {author} {\bibfnamefont {S.}~\bibnamefont {Alcott}} \emph {et~al.},\ }\bibfield  {title} {\enquote {\bibinfo {title} {{Search for millicharged particles in proton-proton collisions at $\sqrt{s} = 13.6$ TeV}},}\ }\href@noop {} {\  (\bibinfo {year} {2025})},\ \Eprint {http://arxiv.org/abs/2506.02251} {arXiv:2506.02251 [hep-ex]} \BibitemShut {NoStop}%
\bibitem [{\citenamefont {Chang}\ \emph {et~al.}(2018)\citenamefont {Chang}, \citenamefont {Essig},\ and\ \citenamefont {McDermott}}]{Chang:2018rso}%
  \BibitemOpen
  \bibfield  {author} {\bibinfo {author} {\bibfnamefont {Jae~Hyeok}\ \bibnamefont {Chang}}, \bibinfo {author} {\bibfnamefont {Rouven}\ \bibnamefont {Essig}}, \ and\ \bibinfo {author} {\bibfnamefont {Samuel~D.}\ \bibnamefont {McDermott}},\ }\bibfield  {title} {\enquote {\bibinfo {title} {{Supernova 1987A Constraints on Sub-GeV Dark Sectors, Millicharged Particles, the QCD Axion, and an Axion-like Particle}},}\ }\href {\doibase 10.1007/JHEP09(2018)051} {\bibfield  {journal} {\bibinfo  {journal} {JHEP}\ }\textbf {\bibinfo {volume} {09}},\ \bibinfo {pages} {051} (\bibinfo {year} {2018})},\ \Eprint {http://arxiv.org/abs/1803.00993} {arXiv:1803.00993 [hep-ph]} \BibitemShut {NoStop}%
\bibitem [{\citenamefont {Budker}\ \emph {et~al.}(2022)\citenamefont {Budker}, \citenamefont {Graham}, \citenamefont {Ramani}, \citenamefont {Schmidt-Kaler}, \citenamefont {Smorra},\ and\ \citenamefont {Ulmer}}]{Budker:2021quh}%
  \BibitemOpen
  \bibfield  {author} {\bibinfo {author} {\bibfnamefont {Dmitry}\ \bibnamefont {Budker}}, \bibinfo {author} {\bibfnamefont {Peter~W.}\ \bibnamefont {Graham}}, \bibinfo {author} {\bibfnamefont {Harikrishnan}\ \bibnamefont {Ramani}}, \bibinfo {author} {\bibfnamefont {Ferdinand}\ \bibnamefont {Schmidt-Kaler}}, \bibinfo {author} {\bibfnamefont {Christian}\ \bibnamefont {Smorra}}, \ and\ \bibinfo {author} {\bibfnamefont {Stefan}\ \bibnamefont {Ulmer}},\ }\bibfield  {title} {\enquote {\bibinfo {title} {{Millicharged Dark Matter Detection with Ion Traps}},}\ }\href {\doibase 10.1103/PRXQuantum.3.010330} {\bibfield  {journal} {\bibinfo  {journal} {PRX Quantum}\ }\textbf {\bibinfo {volume} {3}},\ \bibinfo {pages} {010330} (\bibinfo {year} {2022})},\ \Eprint {http://arxiv.org/abs/2108.05283} {arXiv:2108.05283 [hep-ph]} \BibitemShut {NoStop}%
\bibitem [{\citenamefont {Kim}\ \emph {et~al.}(2007)\citenamefont {Kim}, \citenamefont {Lee}, \citenamefont {Lee}, \citenamefont {Perl}, \citenamefont {Halyo},\ and\ \citenamefont {Loomba}}]{Kim:2007zzs}%
  \BibitemOpen
  \bibfield  {author} {\bibinfo {author} {\bibfnamefont {Peter~C.}\ \bibnamefont {Kim}}, \bibinfo {author} {\bibfnamefont {Eric~R.}\ \bibnamefont {Lee}}, \bibinfo {author} {\bibfnamefont {Irwin~T.}\ \bibnamefont {Lee}}, \bibinfo {author} {\bibfnamefont {Martin~L.}\ \bibnamefont {Perl}}, \bibinfo {author} {\bibfnamefont {Valerie}\ \bibnamefont {Halyo}}, \ and\ \bibinfo {author} {\bibfnamefont {Dinesh}\ \bibnamefont {Loomba}},\ }\bibfield  {title} {\enquote {\bibinfo {title} {{Search for fractional-charge particles in meteoritic material}},}\ }\href {\doibase 10.1103/PhysRevLett.99.161804} {\bibfield  {journal} {\bibinfo  {journal} {Phys. Rev. Lett.}\ }\textbf {\bibinfo {volume} {99}},\ \bibinfo {pages} {161804} (\bibinfo {year} {2007})}\BibitemShut {NoStop}%
\bibitem [{\citenamefont {Moore}\ \emph {et~al.}(2014)\citenamefont {Moore}, \citenamefont {Rider},\ and\ \citenamefont {Gratta}}]{Moore:2014yba}%
  \BibitemOpen
  \bibfield  {author} {\bibinfo {author} {\bibfnamefont {David~C.}\ \bibnamefont {Moore}}, \bibinfo {author} {\bibfnamefont {Alexander~D.}\ \bibnamefont {Rider}}, \ and\ \bibinfo {author} {\bibfnamefont {Giorgio}\ \bibnamefont {Gratta}},\ }\bibfield  {title} {\enquote {\bibinfo {title} {{Search for Millicharged Particles Using Optically Levitated Microspheres}},}\ }\href {\doibase 10.1103/PhysRevLett.113.251801} {\bibfield  {journal} {\bibinfo  {journal} {Phys. Rev. Lett.}\ }\textbf {\bibinfo {volume} {113}},\ \bibinfo {pages} {251801} (\bibinfo {year} {2014})},\ \Eprint {http://arxiv.org/abs/1408.4396} {arXiv:1408.4396 [hep-ex]} \BibitemShut {NoStop}%
\bibitem [{\citenamefont {Afek}\ \emph {et~al.}(2021)\citenamefont {Afek}, \citenamefont {Monteiro}, \citenamefont {Wang}, \citenamefont {Siegel}, \citenamefont {Ghosh},\ and\ \citenamefont {Moore}}]{Afek:2020lek}%
  \BibitemOpen
  \bibfield  {author} {\bibinfo {author} {\bibfnamefont {Gadi}\ \bibnamefont {Afek}}, \bibinfo {author} {\bibfnamefont {Fernando}\ \bibnamefont {Monteiro}}, \bibinfo {author} {\bibfnamefont {Jiaxiang}\ \bibnamefont {Wang}}, \bibinfo {author} {\bibfnamefont {Benjamin}\ \bibnamefont {Siegel}}, \bibinfo {author} {\bibfnamefont {Sumita}\ \bibnamefont {Ghosh}}, \ and\ \bibinfo {author} {\bibfnamefont {David~C.}\ \bibnamefont {Moore}},\ }\bibfield  {title} {\enquote {\bibinfo {title} {{Limits on the abundance of millicharged particles bound to matter}},}\ }\href {\doibase 10.1103/PhysRevD.104.012004} {\bibfield  {journal} {\bibinfo  {journal} {Phys. Rev. D}\ }\textbf {\bibinfo {volume} {104}},\ \bibinfo {pages} {012004} (\bibinfo {year} {2021})},\ \Eprint {http://arxiv.org/abs/2012.08169} {arXiv:2012.08169 [hep-ex]} \BibitemShut {NoStop}%
\bibitem [{\citenamefont {Adari}\ \emph {et~al.}(2025)\citenamefont {Adari} \emph {et~al.}}]{SENSEI:2023zdf}%
  \BibitemOpen
  \bibfield  {author} {\bibinfo {author} {\bibfnamefont {Prakruth}\ \bibnamefont {Adari}} \emph {et~al.} (\bibinfo {collaboration} {SENSEI}),\ }\bibfield  {title} {\enquote {\bibinfo {title} {{First Direct-Detection Results on Sub-GeV Dark Matter Using the SENSEI Detector at SNOLAB}},}\ }\href {\doibase 10.1103/PhysRevLett.134.011804} {\bibfield  {journal} {\bibinfo  {journal} {Phys. Rev. Lett.}\ }\textbf {\bibinfo {volume} {134}},\ \bibinfo {pages} {011804} (\bibinfo {year} {2025})},\ \Eprint {http://arxiv.org/abs/2312.13342} {arXiv:2312.13342 [astro-ph.CO]} \BibitemShut {NoStop}%
\bibitem [{\citenamefont {Aggarwal}\ \emph {et~al.}(2025)\citenamefont {Aggarwal} \emph {et~al.}}]{DAMIC-M:2025luv}%
  \BibitemOpen
  \bibfield  {author} {\bibinfo {author} {\bibfnamefont {K.}~\bibnamefont {Aggarwal}} \emph {et~al.} (\bibinfo {collaboration} {DAMIC-M}),\ }\bibfield  {title} {\enquote {\bibinfo {title} {{Probing Benchmark Models of Hidden-Sector Dark Matter with DAMIC-M}},}\ }\href {\doibase 10.1103/2tcc-bqck} {\bibfield  {journal} {\bibinfo  {journal} {Phys. Rev. Lett.}\ }\textbf {\bibinfo {volume} {135}},\ \bibinfo {pages} {071002} (\bibinfo {year} {2025})},\ \Eprint {http://arxiv.org/abs/2503.14617} {arXiv:2503.14617 [hep-ex]} \BibitemShut {NoStop}%
\bibitem [{\citenamefont {Iles}\ \emph {et~al.}(2025)\citenamefont {Iles}, \citenamefont {Heeba},\ and\ \citenamefont {Schutz}}]{Iles:2024zka}%
  \BibitemOpen
  \bibfield  {author} {\bibinfo {author} {\bibfnamefont {Ella}\ \bibnamefont {Iles}}, \bibinfo {author} {\bibfnamefont {Saniya}\ \bibnamefont {Heeba}}, \ and\ \bibinfo {author} {\bibfnamefont {Katelin}\ \bibnamefont {Schutz}},\ }\bibfield  {title} {\enquote {\bibinfo {title} {{Dark Matter Direct Detection Experiments Are Sensitive to the Millicharged Background}},}\ }\href {\doibase 10.1103/PhysRevLett.134.121002} {\bibfield  {journal} {\bibinfo  {journal} {Phys. Rev. Lett.}\ }\textbf {\bibinfo {volume} {134}},\ \bibinfo {pages} {121002} (\bibinfo {year} {2025})},\ \Eprint {http://arxiv.org/abs/2407.21096} {arXiv:2407.21096 [hep-ph]} \BibitemShut {NoStop}%
\bibitem [{\citenamefont {Pospelov}\ and\ \citenamefont {Ramani}(2021)}]{Pospelov:2020ktu}%
  \BibitemOpen
  \bibfield  {author} {\bibinfo {author} {\bibfnamefont {Maxim}\ \bibnamefont {Pospelov}}\ and\ \bibinfo {author} {\bibfnamefont {Harikrishnan}\ \bibnamefont {Ramani}},\ }\bibfield  {title} {\enquote {\bibinfo {title} {{Earth-bound millicharge relics}},}\ }\href {\doibase 10.1103/PhysRevD.103.115031} {\bibfield  {journal} {\bibinfo  {journal} {Phys. Rev. D}\ }\textbf {\bibinfo {volume} {103}},\ \bibinfo {pages} {115031} (\bibinfo {year} {2021})},\ \Eprint {http://arxiv.org/abs/2012.03957} {arXiv:2012.03957 [hep-ph]} \BibitemShut {NoStop}%
\bibitem [{\citenamefont {Berlin}\ \emph {et~al.}(2024)\citenamefont {Berlin}, \citenamefont {Liu}, \citenamefont {Pospelov},\ and\ \citenamefont {Ramani}}]{Berlin:2023zpn}%
  \BibitemOpen
  \bibfield  {author} {\bibinfo {author} {\bibfnamefont {Asher}\ \bibnamefont {Berlin}}, \bibinfo {author} {\bibfnamefont {Hongwan}\ \bibnamefont {Liu}}, \bibinfo {author} {\bibfnamefont {Maxim}\ \bibnamefont {Pospelov}}, \ and\ \bibinfo {author} {\bibfnamefont {Harikrishnan}\ \bibnamefont {Ramani}},\ }\bibfield  {title} {\enquote {\bibinfo {title} {{Terrestrial density of strongly-coupled relics}},}\ }\href {\doibase 10.1103/PhysRevD.109.075027} {\bibfield  {journal} {\bibinfo  {journal} {Phys. Rev. D}\ }\textbf {\bibinfo {volume} {109}},\ \bibinfo {pages} {075027} (\bibinfo {year} {2024})},\ \Eprint {http://arxiv.org/abs/2302.06619} {arXiv:2302.06619 [hep-ph]} \BibitemShut {NoStop}%
\bibitem [{\citenamefont {Plestid}\ \emph {et~al.}(2020)\citenamefont {Plestid}, \citenamefont {Takhistov}, \citenamefont {Tsai}, \citenamefont {Bringmann}, \citenamefont {Kusenko},\ and\ \citenamefont {Pospelov}}]{Plestid:2020kdm}%
  \BibitemOpen
  \bibfield  {author} {\bibinfo {author} {\bibfnamefont {Ryan}\ \bibnamefont {Plestid}}, \bibinfo {author} {\bibfnamefont {Volodymyr}\ \bibnamefont {Takhistov}}, \bibinfo {author} {\bibfnamefont {Yu-Dai}\ \bibnamefont {Tsai}}, \bibinfo {author} {\bibfnamefont {Torsten}\ \bibnamefont {Bringmann}}, \bibinfo {author} {\bibfnamefont {Alexander}\ \bibnamefont {Kusenko}}, \ and\ \bibinfo {author} {\bibfnamefont {Maxim}\ \bibnamefont {Pospelov}},\ }\bibfield  {title} {\enquote {\bibinfo {title} {{New Constraints on Millicharged Particles from Cosmic-ray Production}},}\ }\href {\doibase 10.1103/PhysRevD.102.115032} {\bibfield  {journal} {\bibinfo  {journal} {Phys. Rev. D}\ }\textbf {\bibinfo {volume} {102}},\ \bibinfo {pages} {115032} (\bibinfo {year} {2020})},\ \Eprint {http://arxiv.org/abs/2002.11732} {arXiv:2002.11732 [hep-ph]} \BibitemShut {NoStop}%
\bibitem [{\citenamefont {Gao}\ and\ \citenamefont {Pospelov}(2025)}]{Gao:2025ykc}%
  \BibitemOpen
  \bibfield  {author} {\bibinfo {author} {\bibfnamefont {Ting}\ \bibnamefont {Gao}}\ and\ \bibinfo {author} {\bibfnamefont {Maxim}\ \bibnamefont {Pospelov}},\ }\bibfield  {title} {\enquote {\bibinfo {title} {{Constraints on millicharged particles from nuclear gamma-decays}},}\ }\href@noop {} {\  (\bibinfo {year} {2025})},\ \Eprint {http://arxiv.org/abs/2507.17955} {arXiv:2507.17955 [hep-ph]} \BibitemShut {NoStop}%
\bibitem [{\citenamefont {Jaeckel}(2009)}]{Jaeckel:2009dh}%
  \BibitemOpen
  \bibfield  {author} {\bibinfo {author} {\bibfnamefont {Joerg}\ \bibnamefont {Jaeckel}},\ }\bibfield  {title} {\enquote {\bibinfo {title} {{Probing Minicharged Particles with Tests of Coulomb's Law}},}\ }\href {\doibase 10.1103/PhysRevLett.103.080402} {\bibfield  {journal} {\bibinfo  {journal} {Phys. Rev. Lett.}\ }\textbf {\bibinfo {volume} {103}},\ \bibinfo {pages} {080402} (\bibinfo {year} {2009})},\ \Eprint {http://arxiv.org/abs/0904.1547} {arXiv:0904.1547 [hep-ph]} \BibitemShut {NoStop}%
\bibitem [{\citenamefont {Berlin}\ \emph {et~al.}(2025)\citenamefont {Berlin}, \citenamefont {Bogorad}, \citenamefont {W.~Graham},\ and\ \citenamefont {Ramani}}]{forthcoming}%
  \BibitemOpen
  \bibfield  {author} {\bibinfo {author} {\bibfnamefont {Asher}\ \bibnamefont {Berlin}}, \bibinfo {author} {\bibfnamefont {Zachary}\ \bibnamefont {Bogorad}}, \bibinfo {author} {\bibfnamefont {Peter}\ \bibnamefont {W.~Graham}}, \ and\ \bibinfo {author} {\bibfnamefont {Harikrishnan}\ \bibnamefont {Ramani}},\ }\bibfield  {title} {\enquote {\bibinfo {title} {{Electric Accumulation of Millicharged Particles}},}\ }\href@noop {} {\  (\bibinfo {year} {2025})}\BibitemShut {NoStop}%
\bibitem [{\citenamefont {Spavieri}\ \emph {et~al.}(2004)\citenamefont {Spavieri}, \citenamefont {Gillies},\ and\ \citenamefont {Rodriguez}}]{GSpavieri_2004}%
  \BibitemOpen
  \bibfield  {author} {\bibinfo {author} {\bibfnamefont {G}~\bibnamefont {Spavieri}}, \bibinfo {author} {\bibfnamefont {G~T}\ \bibnamefont {Gillies}}, \ and\ \bibinfo {author} {\bibfnamefont {M}~\bibnamefont {Rodriguez}},\ }\bibfield  {title} {\enquote {\bibinfo {title} {Physical implications of coulomb's law},}\ }\href {\doibase 10.1088/0026-1394/41/5/S06} {\bibfield  {journal} {\bibinfo  {journal} {Metrologia}\ }\textbf {\bibinfo {volume} {41}},\ \bibinfo {pages} {S159} (\bibinfo {year} {2004})}\BibitemShut {NoStop}%
\bibitem [{\citenamefont {Tu}\ \emph {et~al.}(2005)\citenamefont {Tu}, \citenamefont {Luo},\ and\ \citenamefont {Gillies}}]{Tu:2005ge}%
  \BibitemOpen
  \bibfield  {author} {\bibinfo {author} {\bibfnamefont {Liang-Cheng}\ \bibnamefont {Tu}}, \bibinfo {author} {\bibfnamefont {Jun}\ \bibnamefont {Luo}}, \ and\ \bibinfo {author} {\bibfnamefont {G.~T.}\ \bibnamefont {Gillies}},\ }\bibfield  {title} {\enquote {\bibinfo {title} {{The mass of the photon}},}\ }\href {\doibase 10.1088/0034-4885/68/1/R02} {\bibfield  {journal} {\bibinfo  {journal} {Rept. Prog. Phys.}\ }\textbf {\bibinfo {volume} {68}},\ \bibinfo {pages} {77--130} (\bibinfo {year} {2005})}\BibitemShut {NoStop}%
\bibitem [{\citenamefont {Maxwell}(1873)}]{maxwell}%
  \BibitemOpen
  \bibfield  {author} {\bibinfo {author} {\bibfnamefont {James~Clerk}\ \bibnamefont {Maxwell}},\ }\href@noop {} {\emph {\bibinfo {title} {A Treatise on Electricity and Magnetism}}}\ (\bibinfo {year} {1873})\BibitemShut {NoStop}%
\bibitem [{\citenamefont {Plimpton}\ and\ \citenamefont {Lawton}(1936)}]{Plimpton:1936ont}%
  \BibitemOpen
  \bibfield  {author} {\bibinfo {author} {\bibfnamefont {S.~J.}\ \bibnamefont {Plimpton}}\ and\ \bibinfo {author} {\bibfnamefont {W.~E.}\ \bibnamefont {Lawton}},\ }\bibfield  {title} {\enquote {\bibinfo {title} {{A Very Accurate Test of Coulomb's Law of Force Between Charges}},}\ }\href {\doibase 10.1103/PhysRev.50.1066} {\bibfield  {journal} {\bibinfo  {journal} {Phys. Rev.}\ }\textbf {\bibinfo {volume} {50}},\ \bibinfo {pages} {1066} (\bibinfo {year} {1936})}\BibitemShut {NoStop}%
\bibitem [{\citenamefont {Bartlett}\ \emph {et~al.}(1970)\citenamefont {Bartlett}, \citenamefont {Goldhagen},\ and\ \citenamefont {Phillips}}]{Bartlett:1970js}%
  \BibitemOpen
  \bibfield  {author} {\bibinfo {author} {\bibfnamefont {D.~F.}\ \bibnamefont {Bartlett}}, \bibinfo {author} {\bibfnamefont {P.~E.}\ \bibnamefont {Goldhagen}}, \ and\ \bibinfo {author} {\bibfnamefont {E.~A.}\ \bibnamefont {Phillips}},\ }\bibfield  {title} {\enquote {\bibinfo {title} {{Experimental Test of Coulomb's Law}},}\ }\href {\doibase 10.1103/PhysRevD.2.483} {\bibfield  {journal} {\bibinfo  {journal} {Phys. Rev. D}\ }\textbf {\bibinfo {volume} {2}},\ \bibinfo {pages} {483--487} (\bibinfo {year} {1970})}\BibitemShut {NoStop}%
\bibitem [{\citenamefont {Cochran}\ and\ \citenamefont {Franken}(1968)}]{Cochran1968}%
  \BibitemOpen
  \bibfield  {author} {\bibinfo {author} {\bibfnamefont {G.~D.}\ \bibnamefont {Cochran}}\ and\ \bibinfo {author} {\bibfnamefont {P.~A.}\ \bibnamefont {Franken}},\ }\href@noop {} {\bibfield  {journal} {\bibinfo  {journal} {Bull. Am. Phys. Soc.}\ }\textbf {\bibinfo {volume} {13}},\ \bibinfo {pages} {1379} (\bibinfo {year} {1968})}\BibitemShut {NoStop}%
\bibitem [{\citenamefont {Williams}\ \emph {et~al.}(1971)\citenamefont {Williams}, \citenamefont {Faller},\ and\ \citenamefont {Hill}}]{Williams:1971ms}%
  \BibitemOpen
  \bibfield  {author} {\bibinfo {author} {\bibfnamefont {E.~R.}\ \bibnamefont {Williams}}, \bibinfo {author} {\bibfnamefont {J.~E.}\ \bibnamefont {Faller}}, \ and\ \bibinfo {author} {\bibfnamefont {H.~A.}\ \bibnamefont {Hill}},\ }\bibfield  {title} {\enquote {\bibinfo {title} {{New experimental test of Coulomb's law: A Laboratory upper limit on the photon rest mass}},}\ }\href {\doibase 10.1103/PhysRevLett.26.721} {\bibfield  {journal} {\bibinfo  {journal} {Phys. Rev. Lett.}\ }\textbf {\bibinfo {volume} {26}},\ \bibinfo {pages} {721--724} (\bibinfo {year} {1971})}\BibitemShut {NoStop}%
\bibitem [{\citenamefont {Crandall}(1983)}]{Crandall:1983ec}%
  \BibitemOpen
  \bibfield  {author} {\bibinfo {author} {\bibfnamefont {R.~E.}\ \bibnamefont {Crandall}},\ }\bibfield  {title} {\enquote {\bibinfo {title} {{PHOTON MASS EXPERIMENT}},}\ }\href {\doibase 10.1119/1.13149} {\bibfield  {journal} {\bibinfo  {journal} {Am. J. Phys.}\ }\textbf {\bibinfo {volume} {51}},\ \bibinfo {pages} {698--702} (\bibinfo {year} {1983})}\BibitemShut {NoStop}%
\bibitem [{\citenamefont {Landau}\ and\ \citenamefont {Lifshitz}(1979)}]{LandauKin}%
  \BibitemOpen
  \bibfield  {author} {\bibinfo {author} {\bibnamefont {Landau}}\ and\ \bibinfo {author} {\bibnamefont {Lifshitz}},\ }\href@noop {} {\emph {\bibinfo {title} {Physical Kinetics: Volume 10}}}\ (\bibinfo {year} {1979})\BibitemShut {NoStop}%
\bibitem [{\citenamefont {Berlin}\ \emph {et~al.}(2020)\citenamefont {Berlin}, \citenamefont {D'Agnolo}, \citenamefont {Ellis}, \citenamefont {Schuster},\ and\ \citenamefont {Toro}}]{Berlin:2019uco}%
  \BibitemOpen
  \bibfield  {author} {\bibinfo {author} {\bibfnamefont {Asher}\ \bibnamefont {Berlin}}, \bibinfo {author} {\bibfnamefont {Raffaele~Tito}\ \bibnamefont {D'Agnolo}}, \bibinfo {author} {\bibfnamefont {Sebastian A.~R.}\ \bibnamefont {Ellis}}, \bibinfo {author} {\bibfnamefont {Philip}\ \bibnamefont {Schuster}}, \ and\ \bibinfo {author} {\bibfnamefont {Natalia}\ \bibnamefont {Toro}},\ }\bibfield  {title} {\enquote {\bibinfo {title} {{Directly Deflecting Particle Dark Matter}},}\ }\href {\doibase 10.1103/PhysRevLett.124.011801} {\bibfield  {journal} {\bibinfo  {journal} {Phys. Rev. Lett.}\ }\textbf {\bibinfo {volume} {124}},\ \bibinfo {pages} {011801} (\bibinfo {year} {2020})},\ \Eprint {http://arxiv.org/abs/1908.06982} {arXiv:1908.06982 [hep-ph]} \BibitemShut {NoStop}%
\bibitem [{\citenamefont {Holdom}(1986)}]{Holdom:1985ag}%
  \BibitemOpen
  \bibfield  {author} {\bibinfo {author} {\bibfnamefont {Bob}\ \bibnamefont {Holdom}},\ }\bibfield  {title} {\enquote {\bibinfo {title} {{Two U(1)'s and Epsilon Charge Shifts}},}\ }\href {\doibase 10.1016/0370-2693(86)91377-8} {\bibfield  {journal} {\bibinfo  {journal} {Phys. Lett. B}\ }\textbf {\bibinfo {volume} {166}},\ \bibinfo {pages} {196--198} (\bibinfo {year} {1986})}\BibitemShut {NoStop}%
\bibitem [{\citenamefont {Carney}\ \emph {et~al.}(2021)\citenamefont {Carney}, \citenamefont {H\"affner}, \citenamefont {Moore},\ and\ \citenamefont {Taylor}}]{Carney:2021irt}%
  \BibitemOpen
  \bibfield  {author} {\bibinfo {author} {\bibfnamefont {Daniel}\ \bibnamefont {Carney}}, \bibinfo {author} {\bibfnamefont {Hartmut}\ \bibnamefont {H\"affner}}, \bibinfo {author} {\bibfnamefont {David~C.}\ \bibnamefont {Moore}}, \ and\ \bibinfo {author} {\bibfnamefont {Jacob~M.}\ \bibnamefont {Taylor}},\ }\bibfield  {title} {\enquote {\bibinfo {title} {{Trapped Electrons and Ions as Particle Detectors}},}\ }\href {\doibase 10.1103/PhysRevLett.127.061804} {\bibfield  {journal} {\bibinfo  {journal} {Phys. Rev. Lett.}\ }\textbf {\bibinfo {volume} {127}},\ \bibinfo {pages} {061804} (\bibinfo {year} {2021})},\ \Eprint {http://arxiv.org/abs/2104.05737} {arXiv:2104.05737 [quant-ph]} \BibitemShut {NoStop}%
\bibitem [{\citenamefont {Berlin}\ \emph {et~al.}(2019)\citenamefont {Berlin}, \citenamefont {Blinov}, \citenamefont {Krnjaic}, \citenamefont {Schuster},\ and\ \citenamefont {Toro}}]{Berlin:2018bsc}%
  \BibitemOpen
  \bibfield  {author} {\bibinfo {author} {\bibfnamefont {Asher}\ \bibnamefont {Berlin}}, \bibinfo {author} {\bibfnamefont {Nikita}\ \bibnamefont {Blinov}}, \bibinfo {author} {\bibfnamefont {Gordan}\ \bibnamefont {Krnjaic}}, \bibinfo {author} {\bibfnamefont {Philip}\ \bibnamefont {Schuster}}, \ and\ \bibinfo {author} {\bibfnamefont {Natalia}\ \bibnamefont {Toro}},\ }\bibfield  {title} {\enquote {\bibinfo {title} {{Dark Matter, Millicharges, Axion and Scalar Particles, Gauge Bosons, and Other New Physics with LDMX}},}\ }\href {\doibase 10.1103/PhysRevD.99.075001} {\bibfield  {journal} {\bibinfo  {journal} {Phys. Rev. D}\ }\textbf {\bibinfo {volume} {99}},\ \bibinfo {pages} {075001} (\bibinfo {year} {2019})},\ \Eprint {http://arxiv.org/abs/1807.01730} {arXiv:1807.01730 [hep-ph]} \BibitemShut {NoStop}%
\bibitem [{\citenamefont {Magill}\ \emph {et~al.}(2019)\citenamefont {Magill}, \citenamefont {Plestid}, \citenamefont {Pospelov},\ and\ \citenamefont {Tsai}}]{Magill:2018tbb}%
  \BibitemOpen
  \bibfield  {author} {\bibinfo {author} {\bibfnamefont {Gabriel}\ \bibnamefont {Magill}}, \bibinfo {author} {\bibfnamefont {Ryan}\ \bibnamefont {Plestid}}, \bibinfo {author} {\bibfnamefont {Maxim}\ \bibnamefont {Pospelov}}, \ and\ \bibinfo {author} {\bibfnamefont {Yu-Dai}\ \bibnamefont {Tsai}},\ }\bibfield  {title} {\enquote {\bibinfo {title} {{Millicharged particles in neutrino experiments}},}\ }\href {\doibase 10.1103/PhysRevLett.122.071801} {\bibfield  {journal} {\bibinfo  {journal} {Phys. Rev. Lett.}\ }\textbf {\bibinfo {volume} {122}},\ \bibinfo {pages} {071801} (\bibinfo {year} {2019})},\ \Eprint {http://arxiv.org/abs/1806.03310} {arXiv:1806.03310 [hep-ph]} \BibitemShut {NoStop}%
\bibitem [{\citenamefont {Harnik}\ \emph {et~al.}(2019)\citenamefont {Harnik}, \citenamefont {Liu},\ and\ \citenamefont {Palamara}}]{Harnik:2019zee}%
  \BibitemOpen
  \bibfield  {author} {\bibinfo {author} {\bibfnamefont {Roni}\ \bibnamefont {Harnik}}, \bibinfo {author} {\bibfnamefont {Zhen}\ \bibnamefont {Liu}}, \ and\ \bibinfo {author} {\bibfnamefont {Ornella}\ \bibnamefont {Palamara}},\ }\bibfield  {title} {\enquote {\bibinfo {title} {{Millicharged Particles in Liquid Argon Neutrino Experiments}},}\ }\href {\doibase 10.1007/JHEP07(2019)170} {\bibfield  {journal} {\bibinfo  {journal} {JHEP}\ }\textbf {\bibinfo {volume} {07}},\ \bibinfo {pages} {170} (\bibinfo {year} {2019})},\ \Eprint {http://arxiv.org/abs/1902.03246} {arXiv:1902.03246 [hep-ph]} \BibitemShut {NoStop}%
\bibitem [{\citenamefont {Perez}\ \emph {et~al.}(2024)\citenamefont {Perez} \emph {et~al.}}]{Oscura:2023qch}%
  \BibitemOpen
  \bibfield  {author} {\bibinfo {author} {\bibfnamefont {Santiago}\ \bibnamefont {Perez}} \emph {et~al.} (\bibinfo {collaboration} {Oscura}),\ }\bibfield  {title} {\enquote {\bibinfo {title} {{Searching for millicharged particles with 1 kg of Skipper-CCDs using the NuMI beam at Fermilab}},}\ }\href {\doibase 10.1007/JHEP02(2024)072} {\bibfield  {journal} {\bibinfo  {journal} {JHEP}\ }\textbf {\bibinfo {volume} {02}},\ \bibinfo {pages} {072} (\bibinfo {year} {2024})},\ \Eprint {http://arxiv.org/abs/2304.08625} {arXiv:2304.08625 [hep-ex]} \BibitemShut {NoStop}%
\bibitem [{\citenamefont {Essig}\ \emph {et~al.}(2024)\citenamefont {Essig}, \citenamefont {Plestid},\ and\ \citenamefont {Singal}}]{Essig:2024ebk}%
  \BibitemOpen
  \bibfield  {author} {\bibinfo {author} {\bibfnamefont {Rouven}\ \bibnamefont {Essig}}, \bibinfo {author} {\bibfnamefont {Ryan}\ \bibnamefont {Plestid}}, \ and\ \bibinfo {author} {\bibfnamefont {Aman}\ \bibnamefont {Singal}},\ }\bibfield  {title} {\enquote {\bibinfo {title} {{Collective excitations and low-energy ionization signatures of relativistic particles in silicon detectors}},}\ }\href {\doibase 10.1038/s42005-024-01904-2} {\bibfield  {journal} {\bibinfo  {journal} {Commun. Phys.}\ }\textbf {\bibinfo {volume} {7}},\ \bibinfo {pages} {416} (\bibinfo {year} {2024})},\ \Eprint {http://arxiv.org/abs/2403.00123} {arXiv:2403.00123 [hep-ph]} \BibitemShut {NoStop}%
\bibitem [{\citenamefont {Mirizzi}\ \emph {et~al.}(2009)\citenamefont {Mirizzi}, \citenamefont {Redondo},\ and\ \citenamefont {Sigl}}]{Mirizzi:2009iz}%
  \BibitemOpen
  \bibfield  {author} {\bibinfo {author} {\bibfnamefont {Alessandro}\ \bibnamefont {Mirizzi}}, \bibinfo {author} {\bibfnamefont {Javier}\ \bibnamefont {Redondo}}, \ and\ \bibinfo {author} {\bibfnamefont {Gunter}\ \bibnamefont {Sigl}},\ }\bibfield  {title} {\enquote {\bibinfo {title} {{Microwave Background Constraints on Mixing of Photons with Hidden Photons}},}\ }\href {\doibase 10.1088/1475-7516/2009/03/026} {\bibfield  {journal} {\bibinfo  {journal} {JCAP}\ }\textbf {\bibinfo {volume} {03}},\ \bibinfo {pages} {026} (\bibinfo {year} {2009})},\ \Eprint {http://arxiv.org/abs/0901.0014} {arXiv:0901.0014 [hep-ph]} \BibitemShut {NoStop}%
\bibitem [{\citenamefont {An}\ \emph {et~al.}(2013)\citenamefont {An}, \citenamefont {Pospelov},\ and\ \citenamefont {Pradler}}]{An:2013yfc}%
  \BibitemOpen
  \bibfield  {author} {\bibinfo {author} {\bibfnamefont {Haipeng}\ \bibnamefont {An}}, \bibinfo {author} {\bibfnamefont {Maxim}\ \bibnamefont {Pospelov}}, \ and\ \bibinfo {author} {\bibfnamefont {Josef}\ \bibnamefont {Pradler}},\ }\bibfield  {title} {\enquote {\bibinfo {title} {{New stellar constraints on dark photons}},}\ }\href {\doibase 10.1016/j.physletb.2013.07.008} {\bibfield  {journal} {\bibinfo  {journal} {Phys. Lett. B}\ }\textbf {\bibinfo {volume} {725}},\ \bibinfo {pages} {190--195} (\bibinfo {year} {2013})},\ \Eprint {http://arxiv.org/abs/1302.3884} {arXiv:1302.3884 [hep-ph]} \BibitemShut {NoStop}%
\end{thebibliography}%

\clearpage
\newpage
\maketitle
\onecolumngrid
\begin{center}
\textbf{\large Cavendish Tests of Terrestrial Millicharged Particles} \\ 
\vspace{0.05in}
{ \it \large Supplemental Material}\\ 
\vspace{0.05in}
{}
{Asher Berlin, Zachary Bogorad, Peter W. Graham, and Harikrishnan Ramani}

\end{center}
\setcounter{equation}{0}
\setcounter{figure}{0}
\setcounter{table}{0}
\setcounter{section}{1}
\renewcommand{\theequation}{S\arabic{equation}}
\renewcommand{\thefigure}{S\arabic{figure}}
\renewcommand{\thetable}{S\arabic{table}}
\interfootnotelinepenalty=10000 


\section*{Debye Screening}

Here, we first derive the general form of the mCP charge density in the ballistic limit where mCPs do not scatter or accumulate within the shell. In this case,  Liouville's theorem states that the mCP phase space density is constant in time. We take the initial phase space distribution to be Maxwell-Boltzmann
\be
f (v_\x) = n_\x \, \bigg( \frac{m_\x}{2  \pi \, T_\x} \bigg)^{3/2} \, e^{-m_\x  v_\x^2 / 2 T_\x}
~,
\ee
normalized to $n_\x = \int d^3 \vv_\x \, f(v_\x)$, where $v_\x$ is the mCP's velocity.
Inside the shell, the electric potential perturbs the mCP velocity to be
\be
v_\x^\p = \sqrt{v_\x^2 \pm 2 e q_\x \, \phi_0 / m_\x}
~,
\ee
where $\pm$ refers to an attractive/repulsive potential. Thus, for an attractive potential, the perturbed number density is
\be
n_+^\p = \int d^3 \vv_\x^\p ~ f(v_\x) = n_+ \, \Bigg[ \, 2 \, \sqrt{\frac{e q_\x \, \phi_0}{\pi T_\x}} + e^{\frac{e q_\x \phi_0}{T_\x}} \, \text{erfc} \Bigg( \sqrt{\frac{e q_\x \, \phi_0}{T_\x}} \, \Bigg) \, \Bigg]
~,
\ee
while, for a repulsive potential, the perturbed density is
\be
n_-^\p = \int d^3 \vv_\x^\p ~ f(v_\x) = n_- \, e^{- \frac{e q_\x \phi_0}{T_\x}}
~.
\ee
Taking the mCP population to be symmetric in charge, $n_+ = n_- = n_\x / 2$, the induced charge density is thus $\rho_\x =  e q_\x (n_+^\p - n_-^\p)$,  which yields
\be
\label{eq:DebyeBallistic}
\rho_\x =  \frac{e q_\x \, n_\x}{2} \, \Bigg(  \, 2 \, \sqrt{\frac{e q_\x \, \phi_0}{\pi T_\x}} + e^{\frac{e q_\x \phi_0}{T_\x}} \, \text{erfc} \Bigg( \sqrt{\frac{e q_\x \, \phi_0}{T_\x}} \, \Bigg) - e^{- \frac{e q_\x \phi_0}{T_\x}} \Bigg)
\simeq
 m_D^2 \, \phi_0 
\times
\begin{cases}
\sqrt{\frac{T_\x}{\pi \, e q_\x \, \phi_0}} & (e q_\x \phi_0 \gg T_\x)
\\
1 & (e q_\x \phi_0 \ll T_\x)
~.
\end{cases}
\ee
Thus, we see in the weak-coupling limit, we reproduce the standard result for Debye screening, as given previously in \Eq{Debye}.

This form of the charge overdensity in the weak-coupling limit also applies to the strongly-collisional regime. For instance, if mCPs rapidly collide with SM particles, they track a local thermal distribution. In local thermal equilibrium, the perturbed density of either species is related to the unperturbed one by a Boltzmann factor,
\be
n_+^\p = \frac{n_\x}{2} \, e^{\frac{e q_\x \phi_0}{T_\x}}
~~,~~
n_-^\p = \frac{n_\x}{2} \, e^{- \frac{e q_\x \phi_0}{T_\x}}
~.
\ee
In the weak-coupling limit, the amplitude of the charge density is then simply
\be
\label{eq:boltzmannfactors}
\rho_\x = \frac{e q_\x \, n_\x}{2} \, \bigg( e^{\frac{e q_\x \phi_0}{T_\x}} - e^{- \frac{e q_\x \phi_0}{T_\x}} \bigg) \simeq m_D^2 \, \phi_0
~.
\ee
Instead, in the case that a single species of mCPs scatters solely with itself, we can use the continuity and Euler fluid equations,
\begin{align}
&\dt n_\x + \grad \cdot (n_\x V_\x) = 0
\nl
&\dt \Vv_\x + (\Vv_\x \cdot \grad) \Vv_\x = \frac{e q_\x \, \Ev_0}{m_\x} - \frac{\grad P_\x}{m_\x n_\x}
~,
\end{align}
where $\Vv_\x$ is the mCP bulk velocity and $P_\x \simeq T_\x n_\x$ the pressure. Expanding $n_\x$ around its background value and working perturbatively in small-coupling, these equations can be combined into a single equation for the induced charge density,
\be
\label{eq:selfcollisional}
\Big( \dt^2 - \frac{T_\x}{m_\x} \, \grad^2 \Big) \, \rho_\x \simeq - \w_p^2 \, \rho_0
~.
\ee
where $\w_p^2 = (e q_\x)^2 n_\x / m_\x$ is square of the mCP plasma frequency, and we used Gauss's Law $\grad \cdot \Ev_0 = \rho_0$, where $\rho_0$ is the SM charge that sources $E_0$. Thus, if $\rho_0$ oscillates at angular frequency $\w_0 = 2 \pi \n_0 \ll v_\text{th} / R_0$, we have $\grad^2 \rho_\x \simeq m_D^2 \, \rho_0$, and, hence, $\rho_\x$ is sourced analogous to $\phi_0$ but with a relative factor of $-m_D^2$. Once again, this corresponds to a charge density that oscillates at frequency $\w_0$ with amplitude $\rho_\x \simeq m_D^2 \, \phi_0$. A similar derivation applies to mCPs that scatter with the SM or a distinct mCP population (distinguished by an opposite sign in charge). Such a derivation is shown in \Eq{rhocoll2} below.

\section*{High-Frequency Suppression}

Here, we derive the form of \Eq{diff}, which states that the mCP signals are suppressed if the voltage oscillation timescale is shorter than the time for mCPs to pass through the experiment. 

In the collisional regime, this can be seen from the fluid equations,
\begin{align}
\label{eq:fluid1}
& \dt \rho_\x + \grad \cdot \jv_\x = 0 
\nl
& \jv_\x \simeq \rho_\x \, \frac{e q_\x \, \Ev}{m_\x \, \Gamma_p} - D_\x \, \grad \rho_\x
~,
\end{align}
where $\rho_\x$ and $\jv_\x$ are the millicharge and millicurrent charge densities, $\Gamma_p$ is the momentum-exchange rate for mCP-atomic collisions, and $D_\x = T_\x / (\Gamma_p \, m_\x)$ is the mCP diffusion coefficient (scattering between positive and negative mCPs can also be incorporated by simply including this scattering rate in the definition of $\Gamma_p$). Above, the first line is simply continuity of mCPs. The second line follows from the Euler equation in the presence of an external electric field $\Ev$. 

Let us now consider a conducting region that is charged, sourcing an oscillating electric field. We wish to calculate the diffusion of mCPs in the interior of the conductor, where the electric field vanishes. For simplicity, we work in a single spatial dimension, denoted by $x$, and take the electric field to be oscillating at angular frequency $\w_0$. \Eq{fluid1} can be solved directly. Dropping the unphysical piece of the solution that grows exponentially at $x \to \infty$, and normalizing $\rho_\x$ at $x = 0$, the solution for $x > 0$ is then
\be
\label{eq:highw1}
\rho_\x (x) = \rho_\x (0)  \, e^{- x / L_D} \, e^{i (\w_0 t - x / L_D)}
~~,~~
L_D \equiv \sqrt{2 D_\x / \w_0}
~,
\ee
where $L_D$ is parametrically the distance diffused by an mCP within an oscillation period. Thus, we see that if the conductor is larger than the distance diffused by the mCP within an oscillation time, the resulting charge density is exponentially suppressed in magnitude, and averages out within the total volume.

We can also see in more detail how this arises perturbatively (still working in the collisional regime). Once again starting from the fluid equations, we have that for a symmetric mCP population $n_+ = n_- = n_\x / 2$,
\begin{align}
& \dt n_\pm + \grad \cdot (n_\pm \Vv_\pm) \simeq \dt n_\pm + \frac{n_\x}{2} \, \grad \cdot \Vv_\pm = 0
\nl
& \Vv_\pm \simeq \pm \frac{e q_\x \, \Ev}{\Gamma_p \, m_\x} - D_\x \, \frac{\grad n_\pm}{n_\pm} \simeq \pm \frac{e q_\x \, \Ev}{\Gamma_p \, m_\x} - 2 D_\x \, \frac{\grad n_\pm}{n_\x}
~,
\end{align}
where $n_\pm$ and $\Vv_\pm$ are the number density and bulk velocity of positively- or negatively-charged particles, respectively, and in the second set of equalities we assumed that this plasma is weakly-perturbed by the electric field. Now, let's take the difference of the two sets of equations, and define $\delta n_\x = n_+ - n_-$ and $\delta \Vv_\x = \Vv_+ - \Vv_-$, which gives
\be
\dt \, \delta n_\x + \frac{n_\x}{2} \, \grad \cdot \delta \Vv_\x \simeq 0
~~,~~
\delta \Vv_\x \simeq \frac{2}{\Gamma_p \, m_\x} \, \bigg( e q_\x \, \Ev - T_\x \, \frac{\grad \, \delta n_\x}{n_\x} \bigg)
~.
\ee
These two equations can be used together to give one equation for $\delta n_\x$ after using $\grad \cdot \Ev = \rho_0 + e q_\x \, \delta n_\x$, where $\rho_0$ is an external SM charge density. Then defining $\rho_\x = e q_\x \, \delta n_\x$, we find
\be
\Big[ \dt - D_\x \, \big( \grad^2 - m_D^2 \big) \Big] \, \rho_\x \simeq - \frac{\w_p^2}{\Gamma_p} \, \rho_0
~.
\ee
In the case of a harmonic SM charge density, $\rho_0 \propto e^{i \w_0 t}$, we have that in the long-time limit (in which case transient behavior is suppressed)
\be
\Big[ \grad^2 - \big( m_D^2 + i \w_0 / D_\x \big) \Big] \, \rho_\x (\xv) \simeq m_D^2 \, \rho_0
~.
\ee
Using the method of Green's function, this gives
\be
\rho_\x (\xv) = - \frac{m_D^2}{4 \pi} \, \int d^3 \xv^\p ~ \frac{e^{-\sqrt{m_D^2 + i \w_0 / D_\x} \, |\xv - \xv^\p|}}{|\xv - \xv^\p|} \, \rho_0 (\xv^\p)
~.
\ee
Now, let us take the SM charge density to be a spherical charged shell, $\rho_0 (\xv) = (\phi_0 / R_0) ~ \delta(r - R_0)$ and ignore mCP backreactions, corresponding to $m_D \to 0$. We find that in the interior of the shell, $r < R_0$,
\be
\label{eq:rhocoll2}
\rho_\x (r) \simeq - m_D^2 \, \phi_0 \times e^{-R_0 / L_D} \times e^{i (\w_0 t - R_0 / L_D)} \times  \frac{\sinh{\big[ (1+i) \, r / L_D\big]}}{(1+i) \, r / L_D}
~.
\ee
Note that the last factor approaches unity in the $\w_0 \to 0$ ($L_D \to \infty$) limit. Thus, we see similar exponential suppression as in \Eq{highw1} in the large $\w_0$ limit. 

Alternatively, we can consider mCPs that are free-streaming, and so cannot be effectively described as a fluid. In the perturbative regime, we can use the results of Ref.~\cite{Berlin:2019uco}, which showed that the induced charge density for non-collisional mCPs is
\be
\rho_\x \simeq - \frac{T_\x}{m_\x} \, m_D^2 \, e^{i \w_0 t} \, \int dv ~ d^3 \xv^\p ~ f(v_\x) ~ \frac{\rho_0 (\xv^\p)}{|\xv - \xv^\p|} \, e^{-i \w_0 |\xv - \xv^\p| / v}
~~,~~
f(v_\x) = \frac{e^{-v_\x^2 / \sigma_v^2}}{\pi^{3/2} \, \sigma_v^3}
~,
\ee
where $f(v_\x)$ is the the velocity distribution with dispersion $\sigma_v$, which is related to the DM temperature by $\sigma_v = \sqrt{2 T_\x / m_\x}$. This integral takes a simple analytic form when evaluated at the origin $r = 0$. Once again using a spherical charged shell for $\rho_0$, we then find
\be
\label{eq:rhoFS}
\rho_\x (0, t) \simeq - m_D^2 \, \phi_0 \, e^{i \w_0 t} \times
\begin{cases}
1 & (\w_0 \to 0)
\\
\frac{-i}{\sqrt{3}} ~ e^{- \frac{3}{4} \, (R_0 / L_\text{th})^{2/3}} ~ \sin\Big[ \frac{3^{3/2}}{4} \, (R_0 / L_\text{th})^{2/3}\Big] & (\w_0 \to \infty)
~,
\end{cases}
\ee
where $L_\text{th} \equiv \sqrt{T_\x / m_\x} / \w_0$ is roughly the distance traveled at thermal velocity within an oscillation time. We once again see exponential suppression in the large frequency limit. To summarize, the exponential scaling of \Eq{diff} is shown in Eqs.~\ref{eq:highw1}, \ref{eq:rhocoll2}, and \ref{eq:rhoFS}. 

\section*{Recast Cavendish Tests}

%
\begin{figure}[t!]
\centering
\includegraphics[width=0.7 \textwidth]{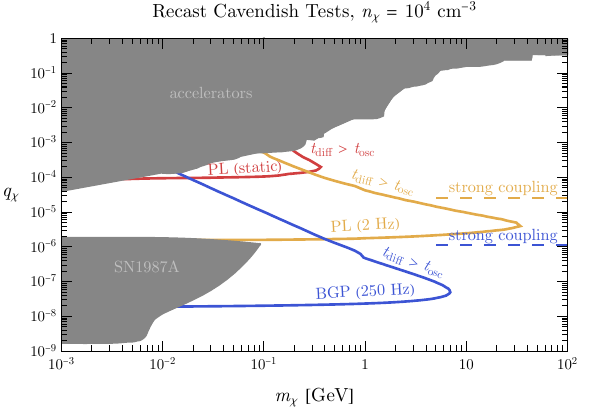}
\caption{Limits recast from the past Cavendish experiments of PL~\cite{Plimpton:1936ont} and BGP~\cite{Bartlett:1970js} on a room-temperature terrestrial population of millicharged particles (solid red, orange, and blue lines), as a function of the particle's mass $m_\x$ and charge $q_\x$ and fixing the ambient density to $n_\x = 10^4 \ \cm^{-3}$. Also shown are previous limits from accelerator probes~\cite{Davidson:2000hf,Haas:2014dda,Prinz:1998ua,ArgoNeuT:2019ckq,milliQan:2021lne,ArguellesDelgado:2021lek,PBC:2025sny,CMS:2024eyx,Alcott:2025rxn} and SN1987A~\cite{Chang:2018rso} (gray). In regions of parameter space labeled ``$t_\text{diff} > t_\text{osc}$," the time for mCPs to diffuse into the experiment is longer than the voltage oscillation time, suppressing the signal exponentially. The point at which $q_\x = 3 T_\x / (e \phi_0)$ is labeled as ``strong coupling," since millicharged particles are able to efficiently electrically bind to the shell for couplings larger than this value.}
\label{fig:CavSummary}
\end{figure}
\begin{figure}[t!]
\centering
\includegraphics[width=0.495 \textwidth]{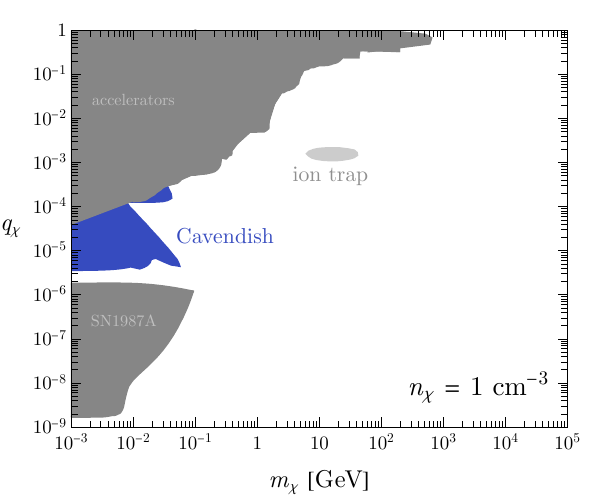}
\includegraphics[width=0.495 \textwidth]{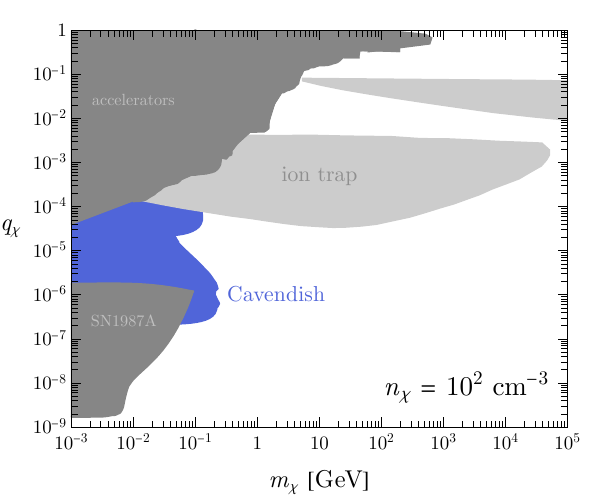}
\includegraphics[width=0.495 \textwidth]{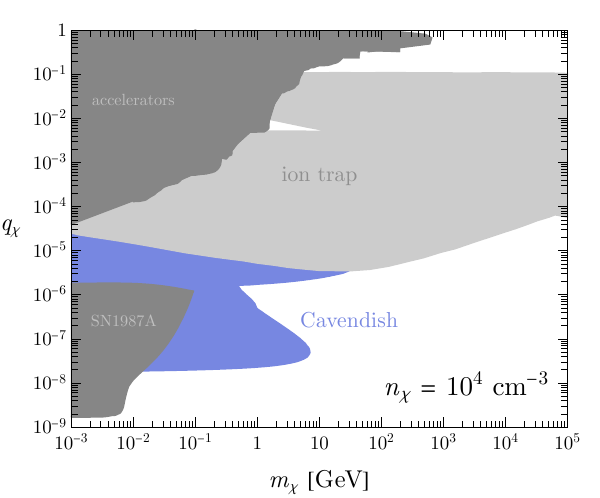}
\includegraphics[width=0.495 \textwidth]{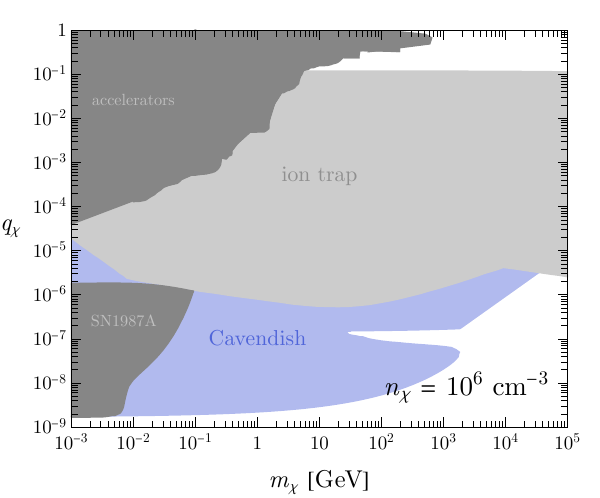}
\caption{As in \Fig{recast}, new limits (blue) on millicharged particles, recast from past Cavendish experiments~\cite{Plimpton:1936ont,Bartlett:1970js}, but now including previous limits derived from ion trap experiments~\cite{Budker:2021quh}. These are shown for the same choices of the terrestrial density $n_\x$ as in \Fig{recast}, except for $n_\x = 0.1 \ \cm^{-3}$ since ion traps are not sensitive to $n_\x < 1 \ \cm^{-3}$.}
\label{fig:ion}
\end{figure}

In \Fig{recast}, we showed the combined limits recast from past Cavendish tests. Here, we present in \Fig{CavSummary} the individual breakdown for each experiment, for a single choice of the terrestrial mCP density $n_\x = 10^4 \ \cm^{-3}$. For sufficiently large couplings, the enhanced interaction strength implies that the time for mCPs to diffuse throughout the experiment is much longer than the voltage oscillation time, suppressing the signal. This region is labeled as ``$t_\text{diff} > t_\text{osc}$." Since the experiments of PL and BGP operated at different frequencies, these regions appears in different parts of parameter space. 

Also, for this representative value of $n_\x = 10^4 \ \cm^{-3}$, these experiments possess sensitivity to sufficiently small couplings for which mCPs do not become electrically bound to the shell, corresponding to $q_\x \lesssim T_\x / (e \phi_0)$. In this case, the signal arises from the weak-coupling limits of Debye screening, as given in \Eq{Debye}. This crossover point at $q_\x \sim T_\x / (e \phi_0)$ is labeled as ``strong coupling" in \Fig{CavSummary}. 

In \Fig{ion}, we compare existing limits from Cavendish tests to recent limits derived from ion trap experiments~\cite{Budker:2021quh}, for various choices of $n_\x$.

\newpage
\section*{Model-Independent Sensitivity}

%
\begin{figure}[t!]
\centering
\includegraphics[width=0.7 \textwidth]{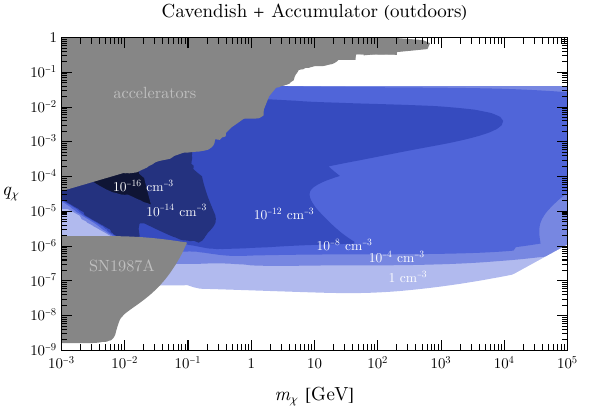}\caption{Analogous to \Fig{recast}, but the projected sensitivity to the terrestrial density $n_\x$ of millicharged particles, for a modified Cavendish test that includes an additional shell fixed at high voltage ($\text{MV}$) and whose interior is operated at high vacuum ($10^{-6} \ \text{atm}$). This additional shell functions as a trap for millicharged particles, enhancing the sensitivity to much smaller ambient densities. Aside from this trap, the experimental setup, including noise levels, is taken to be similar to the previous Cavendish experiment of Ref.~\cite{Bartlett:1970js}.}
\label{fig:future}
\end{figure}

In \Fig{future}, we show the projected sensitivity of our proposed Cavendish test operated within an accumulator shell placed outdoors. In particular, we show the sensitivity for various choices of the ambient mCP density (independent of its origin). Compared to the recast limits from past Cavendish tests in \Fig{recast}, the enhanced reach to larger couplings in \Fig{future} is due to operation in high vacuum, as this decreases the diffusion time throughout the experiment. The sensitivity to charges $q_\x \gg 10^{-2}$ is ultimately limited by the formation of room-temperature mCP-electron bound states as well as the inability of room-temperature mCPs to penetrate the $\sim 10 \ \eV$ surface-barriers at conducting surfaces~\cite{Budker:2021quh,forthcoming}. For the former, we refrain from considering couplings larger than $q_\x \simeq 4 \times 10^{-2}$, while for the latter, we incorporate the fraction of mCPs that are able to overtake this surface-barrier as a simple room-temperature Boltzmann factor.

\end{document}